\documentclass[12pt,preprint]{aastex}

\shorttitle{{\it Spitzer} IRS study of Holmberg~II ULX}
\shortauthors{Berghea et al.}

\begin{document}

\title{The First Detection of [O~IV] from an Ultraluminous X-ray Source with {\it Spitzer}. 
II. Evidence for High Luminosity in Holmberg~II ULX}

\author{C. T. Berghea\altaffilmark{1,2}}
\affil{Physics Department, The Catholic University of America, Washington, DC 20064}
\email{79berghea@cardinalmail.cua.edu}

\author{R. P. Dudik} 
\affil{United States Naval Observatory, Washington, DC 20392}
\email{rpdudik@usno.navy.mil}

\author{K. A. Weaver and T. R. Kallman}
\affil{Laboratory for High Energy Astrophysics, NASA Goddard Space Flight Center, Greenbelt, MD 20771}

\altaffiltext{1}{Computational Physics, Inc., Sprinfield, VA 22151}
\altaffiltext{2}{United States Naval Observatory, Washington, DC 20392;
ciprian.berghea.ctr@usno.navy.mil}

\begin{abstract}

This is the second of two papers examining {\it Spitzer} Infrared Spectrograph (IRS) observations 
of the ultraluminous X-ray source (ULX) in Holmberg~II.  Here we perform detailed photoionization modeling 
of the infrared lines. Our analysis suggests that the luminosity and morphology of the [O~IV] 25.89~$\mu$m emission line
is consistent with photoionization by the soft X-ray and far ultraviolet (FUV) radiation 
from the accretion disk of the binary system and inconsistent with narrow beaming. 
We show that the emission nebula is matter-bounded both in the line of sight direction and to the east, 
and probably radiation-bounded to the west. A bolometric luminosity in excess of 10$^{40}$~erg~s$^{-1}$ would be needed to
produce the measured [O~IV] flux.
We use modeling and previously published studies to conclude that 
shocks likely contribute very little, if at all, to the high-ionization
line fluxes observed in the Holmberg~II ULX.
Additionally, we find that the spectral type of the companion star 
has a surprisingly strong effect on the predicted strength of the [O~IV] emission. 
This finding could explain the origin of [O~IV] in some starburst systems containing black hole binaries.

\end{abstract}


\section{INTRODUCTION}

Despite over two decades of research \citep{fabbiano}, little is known 
about the intrinsic properties of ultraluminous X-ray sources (ULXs).  
They are characterized by extremely high X-ray luminosities, some reaching 10$^{41}$ erg~s$^{-1}$.  
These luminosities cannot be easily explained by normal stellar mass black hole systems, such as those found in our Galaxy. 
Beamed or super-Eddington stellar-mass systems have been proposed as possible models. 
Alternatively, ULXs can harbor intermediate-mass black holes (IMBHs). 
For a complete discussion of various ULX mechanisms, see for example the review papers of 
\citet{millco}, \citet{sor05b}, and \citet{rob07}. Some ULX X-ray spectra show distinct features, 
which have been sometimes interpreted as evidence for the IMBH scenario \citep{mill04}. 
However, the super-Eddington mechanism (maybe together with mild beaming) 
has gained ground in more recent years, backed both by theoretical and observational evidence 
\citep{free06, done06, kun07, king08, sor08, ber08}.

While ULXs  are easy to detect in the X-rays, 
they are often faint or undetected in other energy bands, making them very difficult to characterize.   
The ionization nebulae surrounding some ULXs have become critical for understanding the properties 
of the central source, such as the intrinsic luminosity.  
For instance, \citet{kaa04} (KWZ hereafter) used the optical He~II recombination line ($\lambda$4686) 
to characterize the intrinsic X-ray luminosities of ULXs, as was previously done 
for a ``normal'' X-ray binary by \citet{pak86}. Even [Ne~V] $\lambda$3426, a much higher excitation line 
has been recently detected in a ULX \citep{kaa09}. However, we note that these and analogous studies 
are limited to the optical regime.

The very luminous ULX in the dwarf galaxy Holmberg~II is one such object that has been studied extensively 
in both the X-ray and optical bands \citep{zez99, leh05, goad06, abol07}. It is located inside an ionized nebula (the ``Foot nebula''), 
and shows high ionization optical emission lines coincident with the X-ray source \citep{pak02,abol07}.  
The nebula appears to be perturbed by a moderate velocity field of $\sim$~50~km~s$^{-1}$ \citep{leh05}. 
The optical counterpart of the ULX is a stellar-like bright source located in the ``Heel'' of the nebula, 
having emission consistent with a B supergiant to a main-sequence O star \citep{kaa04}.  
Morever, extended radio emission was detected, approximately coincident with the ionized nebula \citep{mill05}.  
The high ionization recombination line He~II $\lambda$4686 suggests photoionization by the ULX (KWZ), 
and the radio emission is inconsistent with emission from a SNR or an H~II region.

In our first paper \citet{ber09} (Paper~1 hereafter)  we presented  {\it Spitzer} Infrared Spectrograph (IRS) observations 
of the Foot nebula in Holmberg~II.  The low resolution spectral maps presented in that paper 
suggest that the high-ionization [O~IV] line emission is coincident with the ULX position.  
Indeed we found the mid-IR line ratios for this object to be very similar to those seen in standard AGN \citep{gen98, sturm02, dale06}. 
Our preliminary analysis suggested that  [O~IV], like the  He~II recombination line, 
can be used to constrain the spectral energy distribution (SED) of the ULX in both the UV and X-rays, 
thereby constraining the intrinsic X-ray luminosity. To this end, we presented archival X-ray,  
ultraviolet (UV) and infrared (IR) data and constructed a detailed SED of the ULX  
from which photoionization models can be used break the degeneracy in X-ray models 
and extract a reliable bolometric luminosity for the source. 

In this second paper we perform detailed photoionization modeling for the Holmberg~II ULX.
We use the spectral energy distribution (SED) obtained in Paper~1 as the ionizing source, 
and compare the predictions with the observed IR {\it Spitzer} spectrum.
Section~2 presents a summary of the earlier results, which are then used in Section~3
to perform the photoionization modeling.
To better constrain the physical properties, we investigate how the predictions change when we vary the parameters:
the accretion disk temperature, gas density, metallicity, and the stellar companion.
We try both the radiation and matter-bounded geometries to explain the high and low-ionization IR lines.
Finally, in Section~4 we explore shock models as a possible alternative to ionization.
In Sections~5 and 6 we discuss our results and we present our conclusion.

\section{Summary of Observational Results}

\begin{figure}
\epsscale{0.5}
\plotone{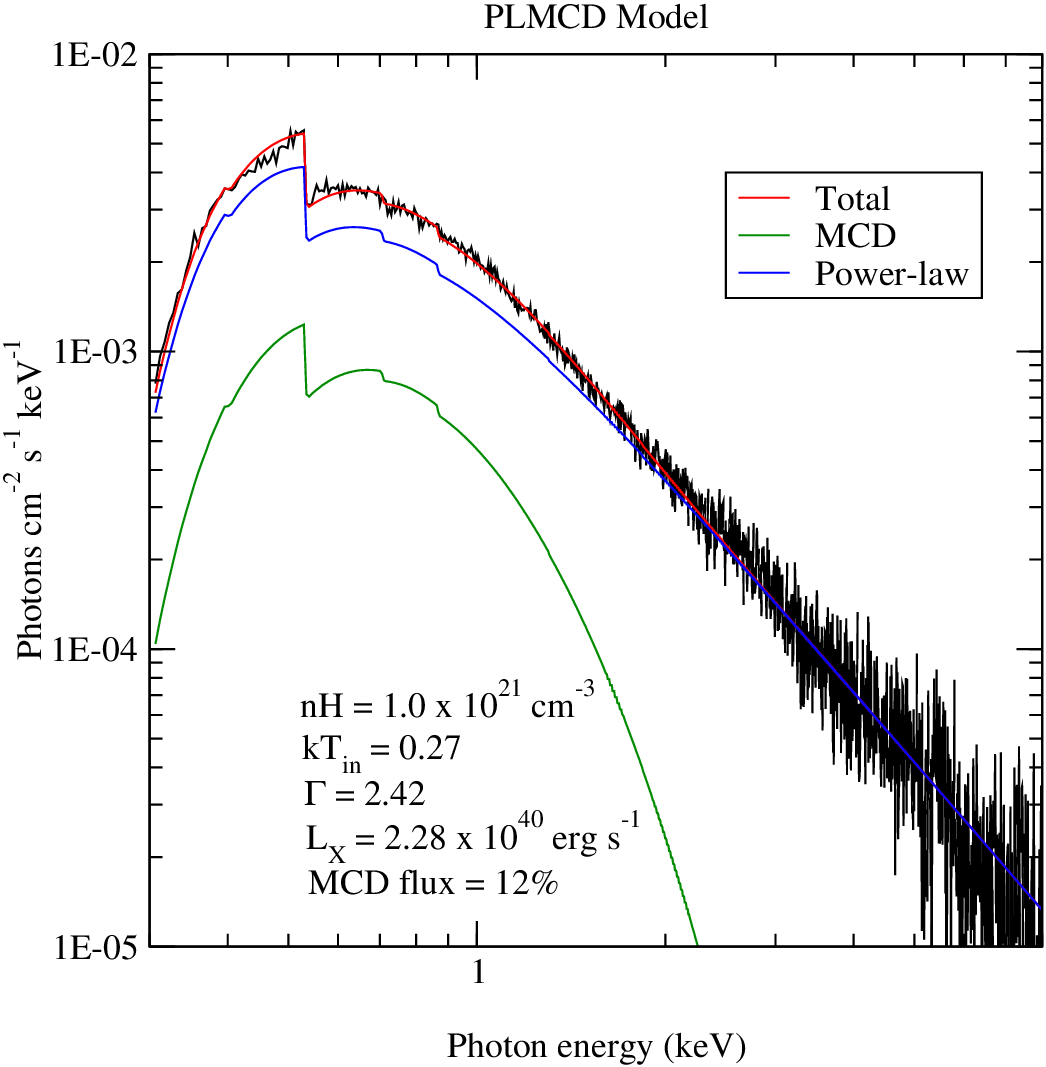}
\caption{
PLMCD model fit to the {\it XMM-Newton} spectrum.
This is the typical model used for ULXs: a multi-color disk component (MCD) plus a power-law at high energies
} 
\label{plmcd}
\end{figure}

\begin{figure}
\epsscale{0.5}
\plotone{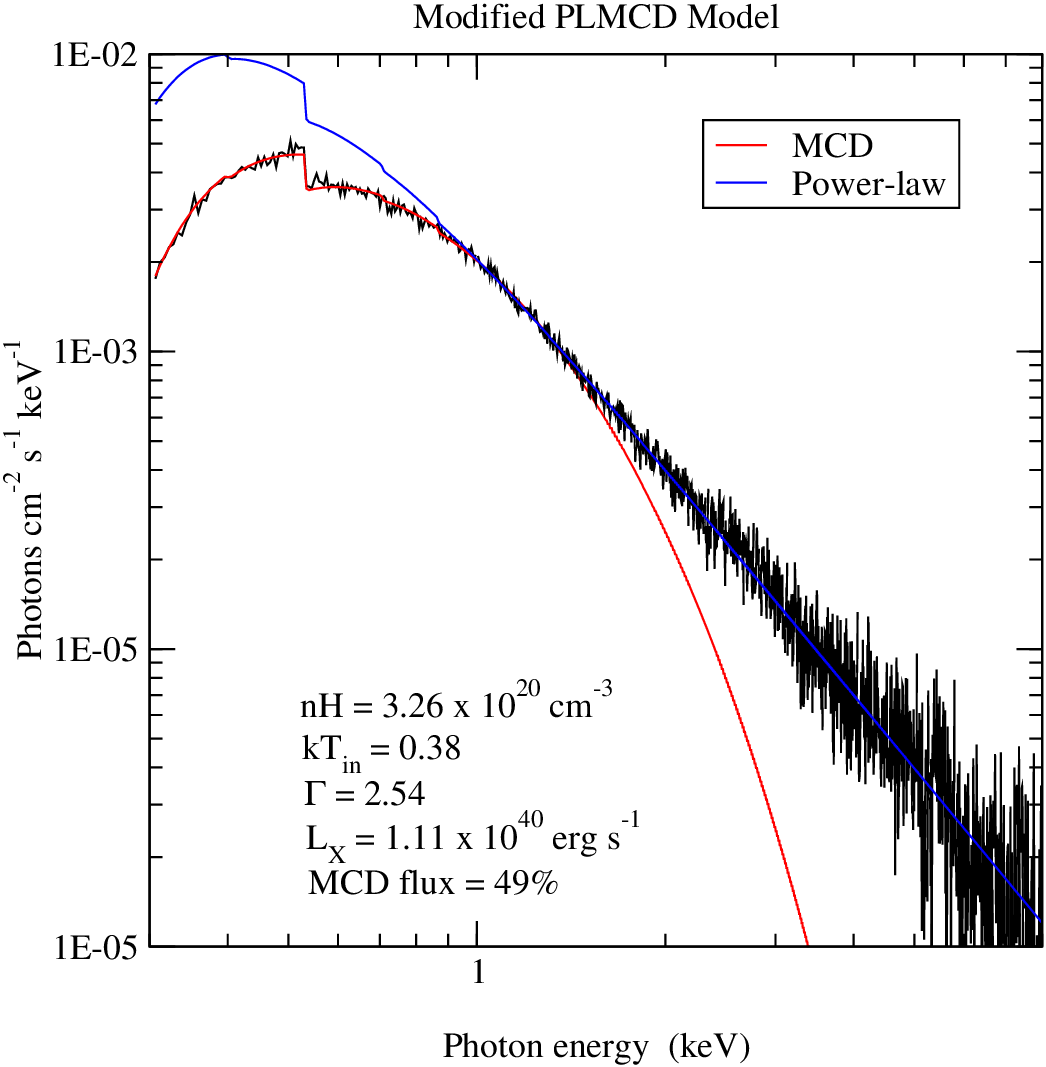}
\caption{
The Modified PLMCD model fit. Here the MCD component and the power-law are fitted separately 
below and above 1~keV, respectively.
} 
\label{Mplmcd}
\end{figure}

\begin{figure}
\epsscale{0.5}
\plotone{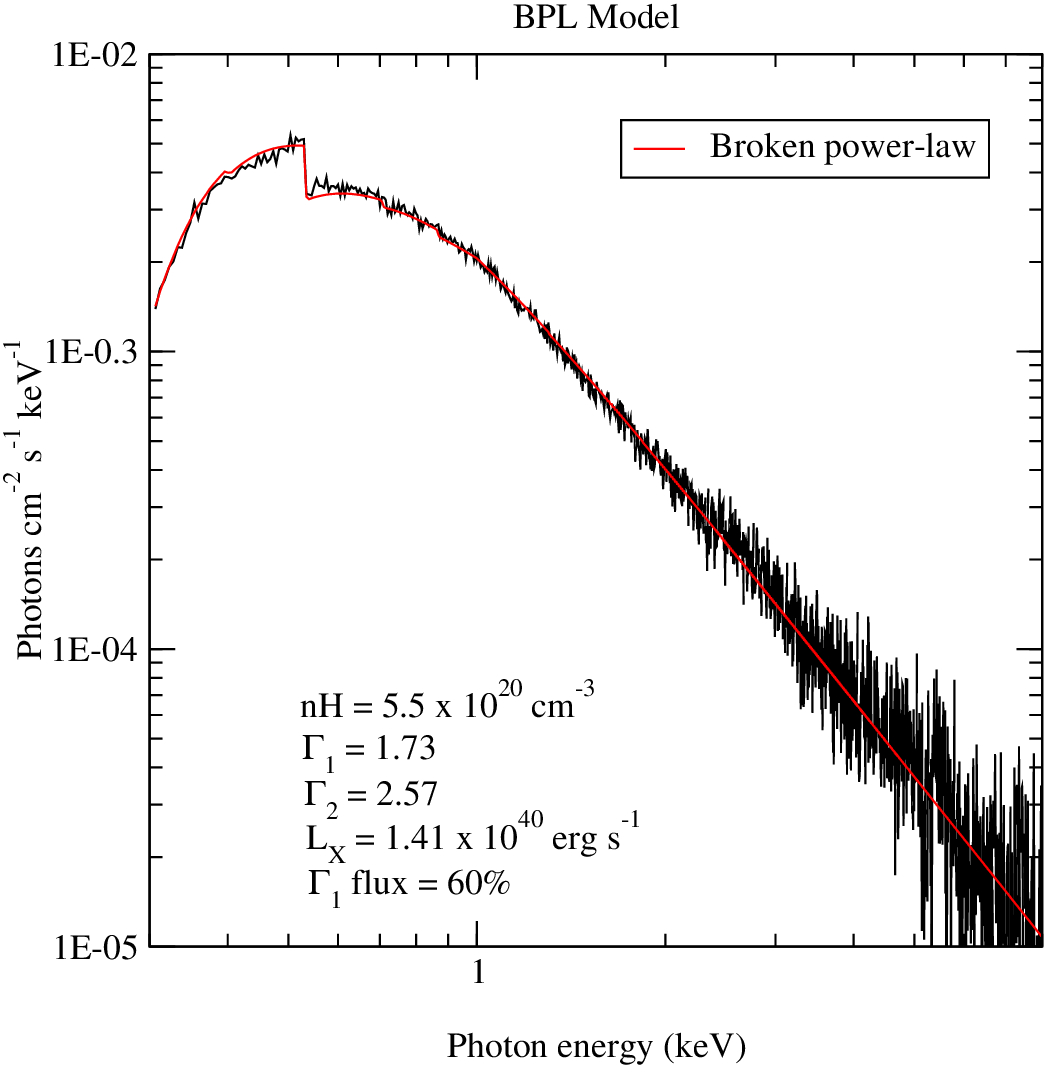}
\caption{
The broken power-law model fit.
The break is at 1~keV, the same energy where the two components are separated in the Modified PLMCD model in Fig.~\ref{Mplmcd}.
} 
\label{bpl}
\end{figure}

The study of ULXs is limited almost exclusively to the X-ray band, where they appear very bright and uncontaminated.  
Studies of these objects at other wavelengths are difficult because the emission 
is generally either undetected or contaminated by circum-stellar or circum-nebular emission, 
making an estimate for the intrinsic bolometric luminosity very difficult.   
This deficiency, especially in the UV, results in degeneracy in X-ray models of the accretion process that cannot be broken.  
Our analysis of the X-ray emission from the ULX in Holmberg~II is an example of such degeneracy.
In Paper~1 three physically distinct models were presented to fit the X-ray spectra for the Holmberg~II ULX.  
The resultant parameters from the fits are given in Table~1. Figures~\ref{plmcd} - \ref{bpl} show the spectra 
and the fitted components for all three models. A brief summary of the three models is as follows:  

\begin{enumerate}

\item {\bf Power-law Plus Multicolor Disk (PLMCD) Model:}  This is a two-component model widely used to describe 
the emission from accreting black holes.  
As the name suggests, the model consists of a multi-color disk component and a power-law component for the hard tail.  
As discussed in Paper~1, when extrapolated to softer energies (the far UV) this model falls apart 
since the power law dominates in the UV. Indeed the model is unrealistic since it is physically impossible 
to have the X-ray photons Componized to lower energies.
This model provides a good fit to the X-ray spectrum and is used extensively in other X-ray studies. 
However we emphasize that it is not physically meaningful \citep[e.g.][]{done06}. 
The PLMCD model is shown here as a base from which modified, but more physically meaningful models are derived. 

\item {\bf The Modified PLMCD Model:} The Modified PLMCD model is also a two-component model, 
but one for which the far-UV extrapolation is from the disk component rather than the power-law component.  
This yields a much more physically meaningful interpretation for the emission.  

\item {\bf Broken Power Law (BPL) Model:} The spectra were also fit with a broken power-law model, 
which is much more flexible when extrapolated to the UV. Such a model would be produced 
if jet emission dominated the full SED. We note that there is currently no direct evidence for jets in ULXs.
It is interesting however that one of the few sources associated with radio emission, 
the famous ULX in NGC~5408, was shown by \citet{kaa03} to be well fitted by a
broken PL \citep[see also][]{sor06, kaa09}.
For more details on our motivation for these models we refer our reader to Paper~1.

\end{enumerate}

\begin{figure}
\epsscale{0.8}
\plotone{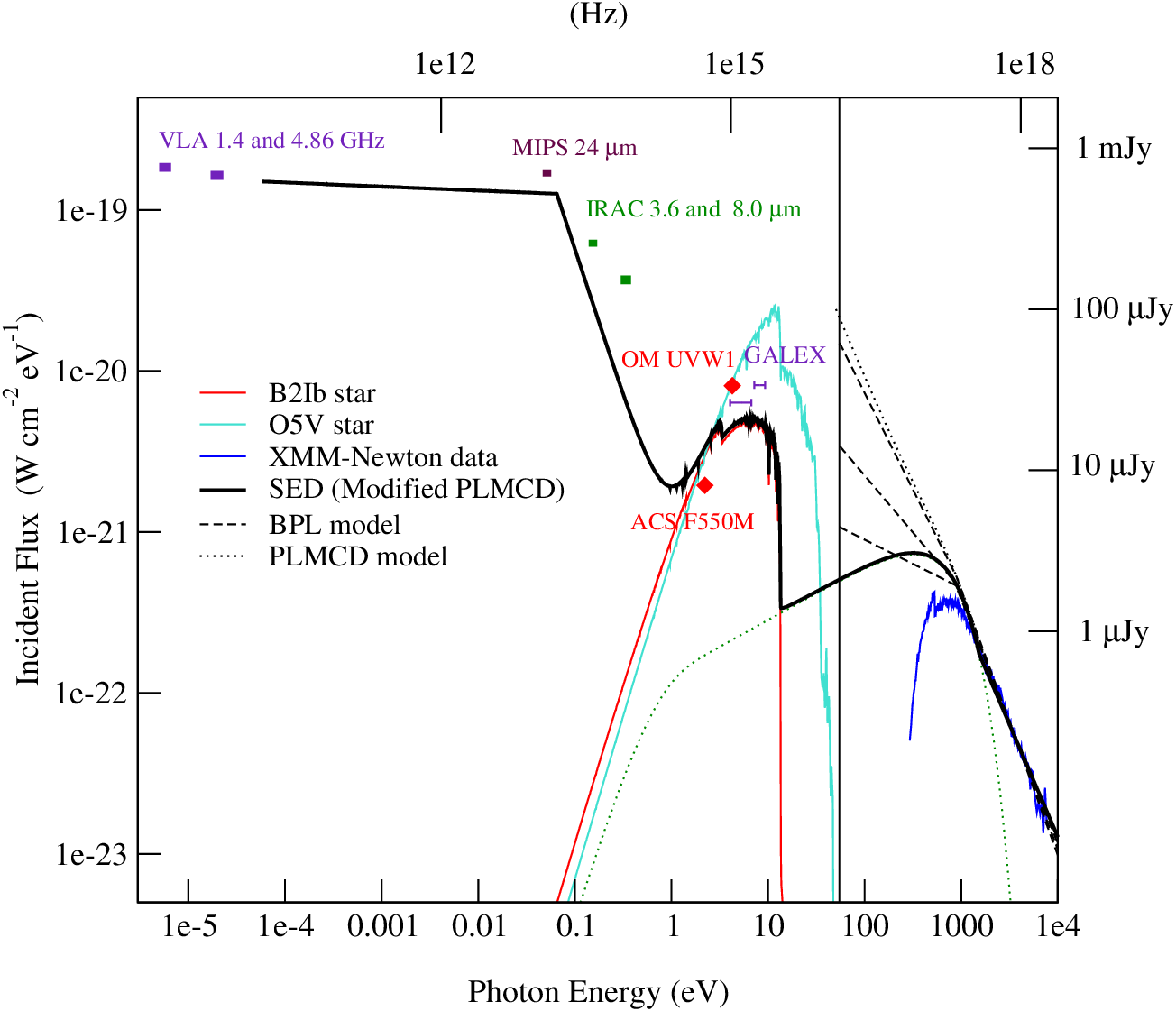}
\caption{
The input SED (Base model) constructed in Paper~1 and used as ionizing source in CLOUDY. 
The radio data is from \citet{mill05}, the V-band magnitude for the optical counterpart of the ULX is quoted from KWZ.
Our measured fluxes in Paper~1 using IRAC, MIPS, OM and {\it GALEX} are upper bounds, and therefore our model 
is below these data points.
The two stellar spectra correspond to the spectral types range consistent with the colors and magnitudes in KWZ.
The three models fit to the X-ray data are shown, extrapolated to the [O~IV] edge at 54.93~eV (shown as a vertical line).
The BPL model was used to obtain a family of broken power-law models, all compatible with the X-ray data within the errors.
We show the lower and upper limits of this family as the lower and upper dashed lines, 
corresponding to $\Gamma=$~1.3 and $\Gamma=$~2.2 in the low energy slope, respectively. 
The middle dashed line represents the best fit BPL model shown in Table~1.
The SED is constructed using the prefered X-ray model, Modified PLMCD.
} \label{sed}
\end{figure}

As can be seen from Table~1, all three models provide reasonably good fits to the X-ray data.  
This degeneracy can only be broken through comprehensive knowledge of the shape of the UV continuum 
and detailed photoionization modeling of high-excitation lines.  In the former case, 
the three X-ray models can be extrapolated to the UV and then compared with observational data.  
The best-fit extrapolated model to the UV spectrum provides some indication of the appropriate physical accretion scenario 
for the observational data. However, contamination to the UV emission by circum-stellar or circum-nebular continuum 
precludes a pure fit. Thus, in addition to this photometric data,
high-ionization spectral lines, 
which require high-energy UV photons to ionize, can be used in combination with photoionization modeling 
to check the robustness of the fit. In Paper~1 we also presented multi-wavelength photometric data 
from archival {\it Spitzer, XMM-Newton} and {\it GALEX} observations of the Holmberg~II ULX, 
and also {\it Hubble Space Telescope} and {\it Very Large Array} photometric data from the literature.  
These data were used to construct a detailed spectral energy distribution (SED) of the ULX (Fig.~\ref{sed}).  
All three X-ray models enumerated above are shown in the plot as dotted, thick solid, and dashed black lines respectively.  
The thick solid black line in Figure~\ref{sed} represents the "base SED model" which is the input spectrum 
for the preliminary photoionization modeling explored below. As discussed in Paper~1 and shown in Figure~\ref{sed}, 
the Modified PLMCD model is chosen as the X-ray model for the base model, 
since it is the only extrapolated model that fits the optical and UV data well. 
Tables~1 and 2 provide the X-ray spectral fitting results and the IR line measurements from the first paper 
that will be the basis of the modeling in the current publication.

\section{CLOUDY Modeling of the IR lines}

To perform detailed photoionization modeling with 
CLOUDY\footnote{ Photoionization modeling was performed using version 07.02.01 of CLOUDY, described by Ferland et al. 1998.},
a number of input parameters need to be defined. Specifically, the geometry of the cloud needs to be described, 
the density and metallicity of the gas needs to be specified, and an input SED needs to be established.  
The following outlines the specific input parameters used for the preliminary CLOUDY run.  
These parameters are later varied to check the final results.   

\begin{enumerate}

\item {\bf Gas Cloud Geometry:}  The input geometry must describe the shape and thickness of the gas cloud.  
Shape includes, for example, a spherical, cylindrical, or plane-parallel geometry. 
The cloud thickness defines a radiation or matter bounded geometry. In their photoionization modeling, KWZ assumed a spherical geometry 
based on resolved optical lines, and assumed the nebula is radiation-bounded. However, they noted that 
the lower-ionization line [O~I] is only detected west of the ULX, thereby suggesting 
a matter-bounded nebula in other directions.  Though we also detected low-ionization lines in the IR spectrum 
(Table~2 in this paper and Figs.~1 and 2 in Paper~1), we begin with the spherical, radiation-bounded geometry 
in an attempt to reproduce the high-excitation [O~IV] emission. We assume the inner radius of the cloud to be 0.1~pc, 
thus maximizing the ionizing flux. To constrain the size of the ionized nebula for the radiation bounded geometry, 
CLOUDY was run down to a temperature of 4000~K. At this temperature, the cloud thickness is 140~pc, 
and the column density is 4.2~$\times$10$^{21}$~cm$^{-2}$, more than ten times the X-ray estimated value.
 
\item {\bf Density:}  Density can be specified as varying, uniform or clumpy. 
In the preliminary CLOUDY model, we assume a constant gas density of 10~cm$^{-3}$ and a filling factor of 1 as per KWZ.
This density (and filling factor) was estimated by KWZ using the optical line surface brightness profiles predicted by CLOUDY.

\item {\bf Metallicity:}  Metallicity is expressed in terms of solar metallicities. 
We chose an input metallicity of 0.1~Z$_{\odot}$, close to that of KWZ and \citet{pil04}. 
 
\item {\bf SED:}  The input SED must specify the companion star and the X-ray model used to fit the UV to X-ray continuum.  
The input SED used in our preliminary analysis is described in Section~2 and results from the multi-wavelength photometric 
and spectroscopic observations presented in Paper~1. This so called "base SED model" utilizes the Modified PLMCD model, 
extrapolated to the UV. For the input SED we used
a B2Ib supergiant spectrum to represent the stellar companion. 
The B2Ib supergiant was chosen because it better fits our UV data than the O5V 
(both are limits to the stellar spectral range suggested by KWZ).  
The bolometric luminosity of this model is 1.34~$\times$10$^{40}$ erg~s$^{-1}$.

\end{enumerate}

\subsection{Results from the Base Model}

\begin{figure}
\epsscale{1.0}
\plotone{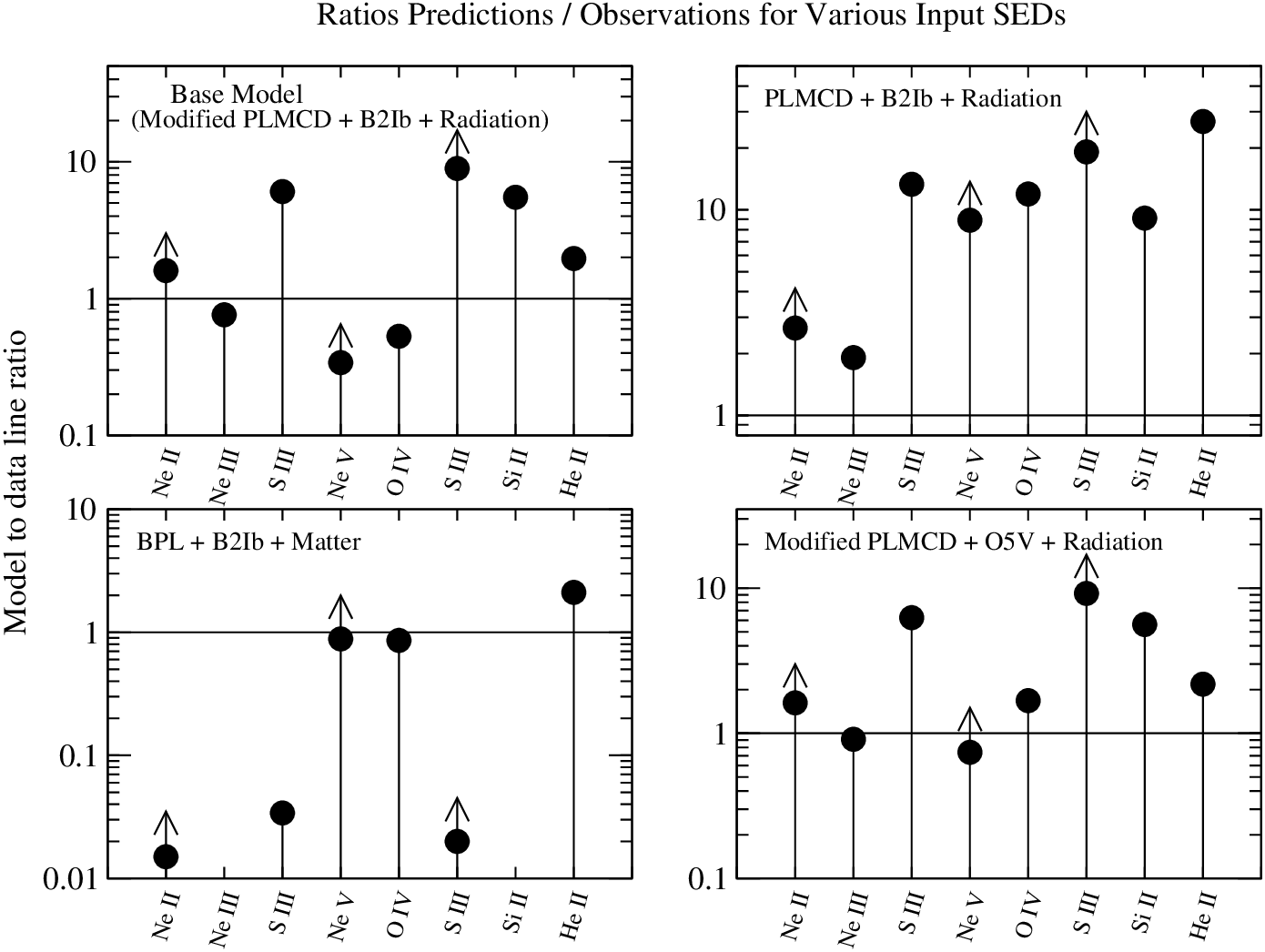}
\caption{
We show predictions for the IR lines from the CLOUDY modeling for various ionizing SEDs.
We also included the He~II $\lambda$4686 line measured by KWZ (2.7~$\times$~10$^{36}$ erg~s$^{-1}$).
They are presented as model-to-data flux ratios. For non-detected lines, the plotted values are lower limits 
and are marked with an arrow pointed upward. 
The predictions for the [Ne~III] and [Si~II] lines are $<$~0.01 in the BPL model (lower-left) and are not shown.
The perfect match line (ratio~$=$~1) is shown on all plots.
The errors are small, similar in size to the data points, therefore are not shown.
The measured line fluxes are presented in the Table~2.
} \label{predictions}
\end{figure}

The results of the first CLOUDY run, with the base models as the ionizing source, are given in Figure~\ref{predictions} 
(top-left panel), as the ratio of the predictions to the observed values. The CLOUDY model 
predicted values for the mid-IR lines are close to the observed values for the high ionization lines.  
A simple change to the base model such as a slightly younger companion star would be enough 
to fit the [O~IV] line observed, as the analysis below indicates.
We also show predictions for the He~II flux measured by KWZ (2.7~$\times$~10$^{36}$ erg~s$^{-1}$).

The low ionization lines are the most discrepant. This disparity between the model and the observations
is likely caused by a matter-bounded geometry in at least some directions of the nebula,
as already suggested by KWZ.
However, many of the model parameters are not well known 
and we explore how changes to these can effect the line-flux predictions in the sections to follow.

We note that sensitivity and aperture effects could explain the discrepancy 
between the low ionization line fluxes predicted by the radiation-bounded base SED model and the observed values.  
Using the CLOUDY output we calculated surface brightness profiles expected for the mid-IR lines. 
CLOUDY predicts a spatial extent of 20 to 25~pc for the He~II which is consistent with the findings of KWZ (15$-$26~pc). 
The [O~IV] emission from the ULX should be extended to a radius of about $\sim$20~pc according to the simulation.   
However, due to the poor resolution of our spectral maps (one LH pixel $=$ 67.5~pc), 
it is unclear whether the model accurately predicts the spatial extent of the [O~IV].  
In addition, the CLOUDY models suggest that while sensitivity may have some effect 
on the observed low-ionization line fluxes, it is not solely responsible for the
high predicted-to-observed line ratios
for the radiation bounded base SED model.

\subsection{Varying the X-ray Source Spectrum to the BPL and PLMCD models}  

We next tried changing the X-ray part of the source spectrum. We first replace the Modified PLMCD model
with the PLMCD model described in Section~2, leaving the other CLOUDY parameters the same. 
The model is extrapolated to the edge of the [O~IV] line where we introduce a break, 
so that it does not dominate the optical flux from the B2 star. 
Results from this CLOUDY simulation are shown in upper-right panel of Figure~\ref{predictions}. 
The values predicted by this model are clearly not consistent with the measured values.  
This model predicts a strong [Ne~V] line, which is not detected in our {\it Spitzer} observations, 
and a He~II luminosity 27 times larger than measured by KWZ. The [O~IV] flux predicted from this model 
is also an order of magnitude larger than the observed flux. We therefore rule out this model 
as a good estimate for the X-ray source spectrum. We note that the PLMCD model has been used 
by many researchers to estimate disk temperatures and even black hole masses. 
Our findings imply that it is likely not an appropriate model for ULX high-energy spectra.  

We then tried replacing the Modified PLMCD model with the BPL model described in Section~2.  
Here we found results similar to the PLMCD model, with [Ne~V] predicted to be detected,
and all the observed lines well overpredicted. We therefore conclude that neither the PLMCD model nor the BPL model
with a radiation bounded geometry provide a good fit to the observed IR spectra. 
This was expected because both these models predict a very strong UV ionizing flux.

\subsection{Varying Other Parameters}

We investigate here in more detail how the strong radiation from the accretion disk photoionizes 
the surrounding ISM and produces the observed lines. Starting again with the base SED model 
in the radiation bounded geometry, we vary the input parameters to see how they affect the emitted spectrum. 
We first examine the dependence of the simulated line luminosities on the X-ray spectrum 
by changing the MCD disk temperature (kT$_{in}$), then the gas density, metallicity and the inner radius of the cloud. 
Finally, we explore how the SED of the stellar companion affects the line luminosities. 
The models presented are based on the multi-wavelength data of the Holmberg~II ULX 
and the known properties of the surrounding ISM, but this analysis can be applied to ULXs in general.

\begin{figure}
\epsscale{0.9}
\plottwo{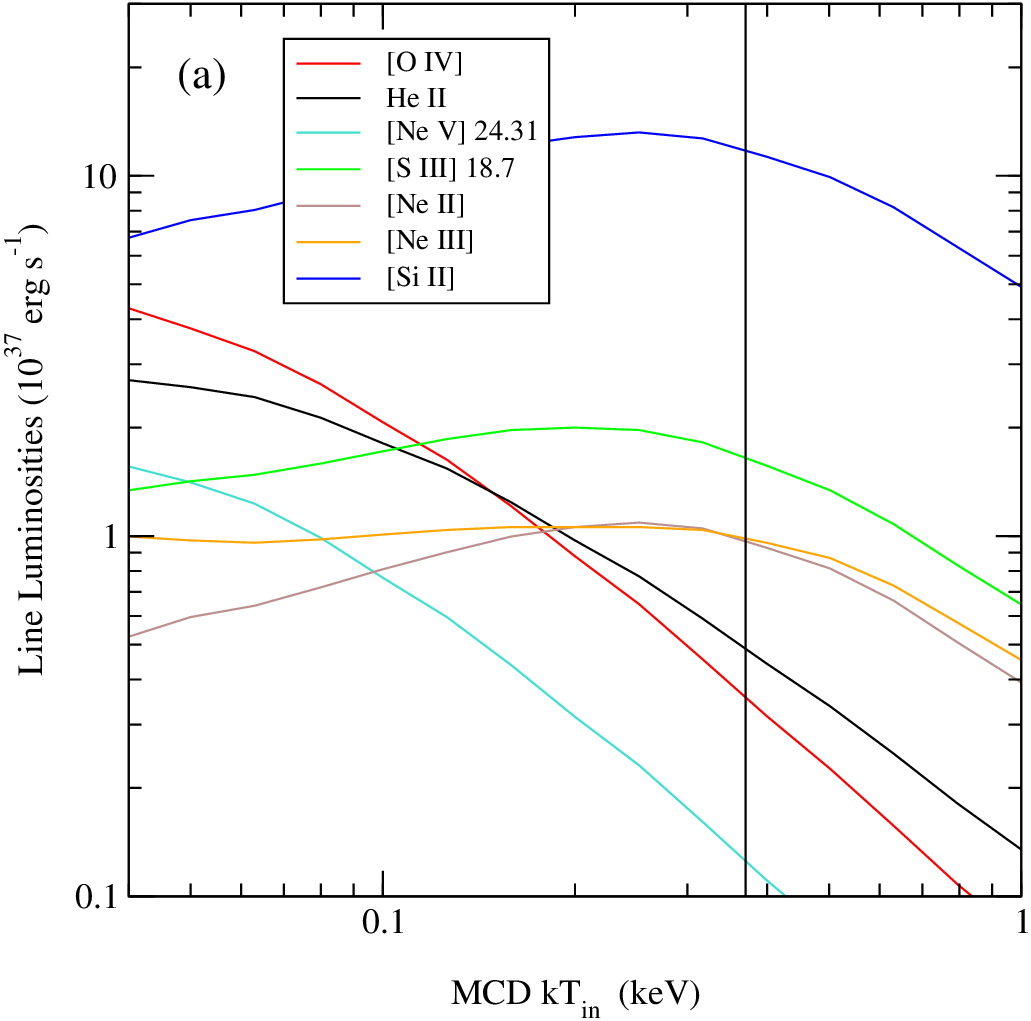}{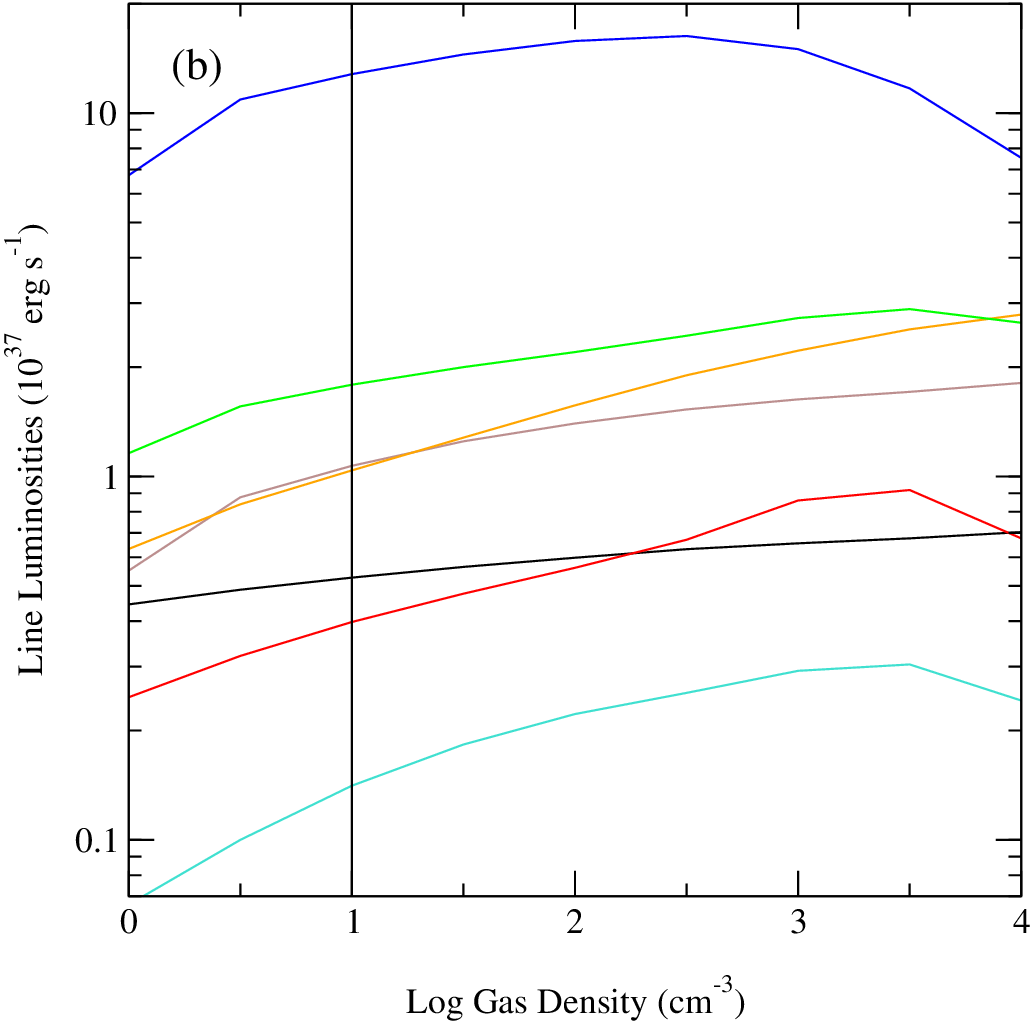}
\caption{
The dependence of the simulated lines from CLOUDY on:
a) MCD inner disk temperature. 
b) Gas density of the cloud.
The vertical lines mark the base model values: kT$_{in}=~$0.38~keV, density~$=$~10~cm$^{-3}$.
The line colors are the same but only labeled in the left panel.
} \label{param1}
\end{figure}

{\bf Varying the Disk Temperature:}  The disk temperature is first altered by shifting the X-ray base SED model 
to lower photon energies, such that the disk luminosity remains constant across the band.  
As can be seen from Figure~\ref{param1}a, when the disk component peaks at lower energies, 
the mid-IR emission lines become stronger. For cooler disks (0.1~keV) the [Ne~V] line should be detected in our observations. 
In contrast, the lower ionization lines are not dependent on the disk temperature over the range 
usually found in ULXs (0.1$-$0.5 keV). We conclude from this that varying the disk temperature 
will not yield better results.

{\bf Varying the Gas Density of the Cloud:}  The results obtained by varying the density of the cloud 
are plotted in Figure~\ref{param1}b. The line luminosities are not particularly sensitive 
to gas density (see also KWZ), though the [O~IV] luminosity increases slightly as the gas approaches the critical density. 
KWZ found that the He~II line is clearly extended to 15$-$25~pc and consistent with a density of $\sim$10~cm$^{-3}$. 
We therefore conclude that varying the density of the gas will not do much to improve the model. 

\begin{figure}
\epsscale{0.9}
\plottwo{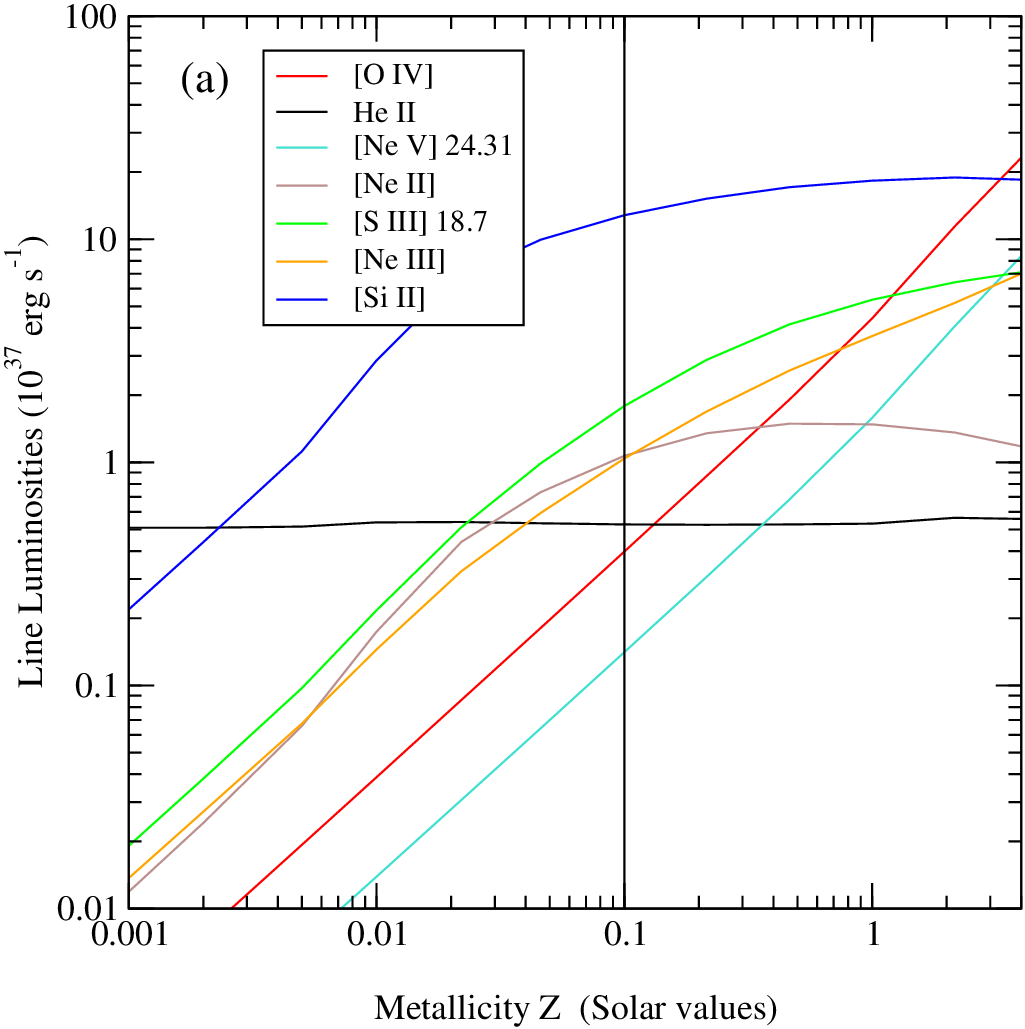}{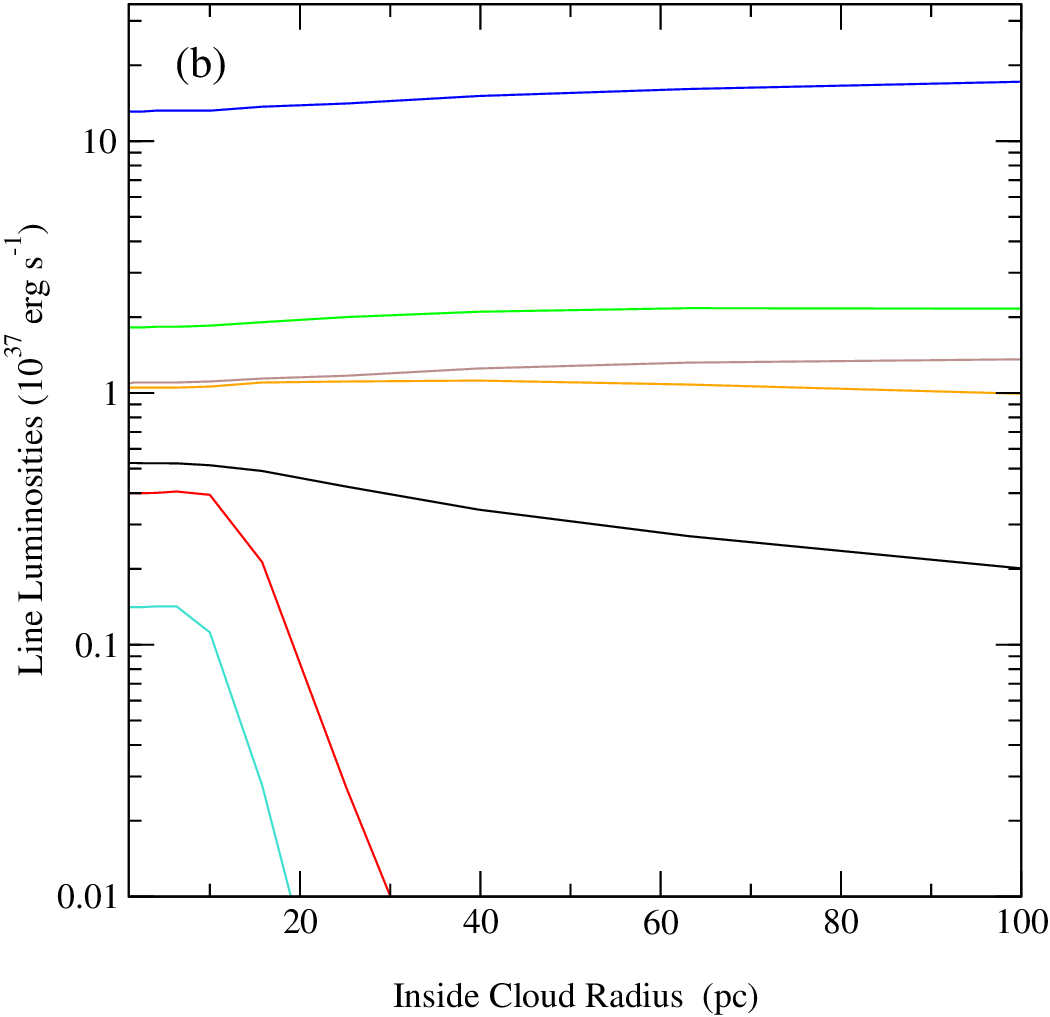}
\caption{
The dependence of the simulated lines from CLOUDY on:
a) Cloud metallicity.
b) The inner radius of the cloud.
The vertical line marks the base model value of Z~$=$~0.1~Z$_\odot$.
The line colors are the same but only labeled in the left panel.
} \label{param2}
\end{figure}

{\bf Varying the Metallicity:} The dependence of the line luminosities on metallicity is shown in Figure~\ref{param2}a.  
The [O~IV] and [Ne~V] line dependence is linear. Interestingly, the base model with a higher input metallicity 
reproduces the measured [O~IV] luminosity very well. In fact a metallicity of 0.19~Z$_\odot$ 
(oxygen abundance of 9.3$\times$10$^{-5}$ in absolute value) reproduces the [O~IV] 
to within the calibration error of the instrument, but without increasing the He~II, 
which is already over-predicted in the current version of the model. 
We conclude from this that the input metallicity may be lower than it should be \citep[see][]{pil04}. 
However, we also note that a higher metallicity will predict higher fluxes for the low ionization lines, 
resulting in an even greater discrepancy for these lines. 
We also explored the effects of gas phase depletion on metallicity and found that depletion 
had very little effect on either the metallicity or the resultant luminosity of the [O~IV].

{\bf Varying the Inner Radius of the Ionizing Cloud:} Figure~\ref{param2}b shows that the ionizing power of the incident flux 
is diluted if the spherical cloud is too far away from the source. If the inner radius is larger than about 10$-$20~pc, 
the luminosities of the [O~IV] and [Ne~V] lines drop dramatically. Interestingly, the He~II line, 
though similar to [O~IV] in ionization potential, is nearly insensitive to the inner cloud radius. 
This is likely because the [O~IV] and [Ne~V] are in a higher stage of excitation than He~II 
and as a result are much more affected by dilution of the ionizing photons 
(in other words a drop in the ionization parameter) from increasing the inner radius.  
We conclude from this that increasing the inner radius will only worsen our original results 
and will not affect the strength of the low ionization lines. Thus we conclude that the inner radius 
is likely close to the source as our original input parameter supposes.

\begin{figure}
\epsscale{0.9}
\plottwo{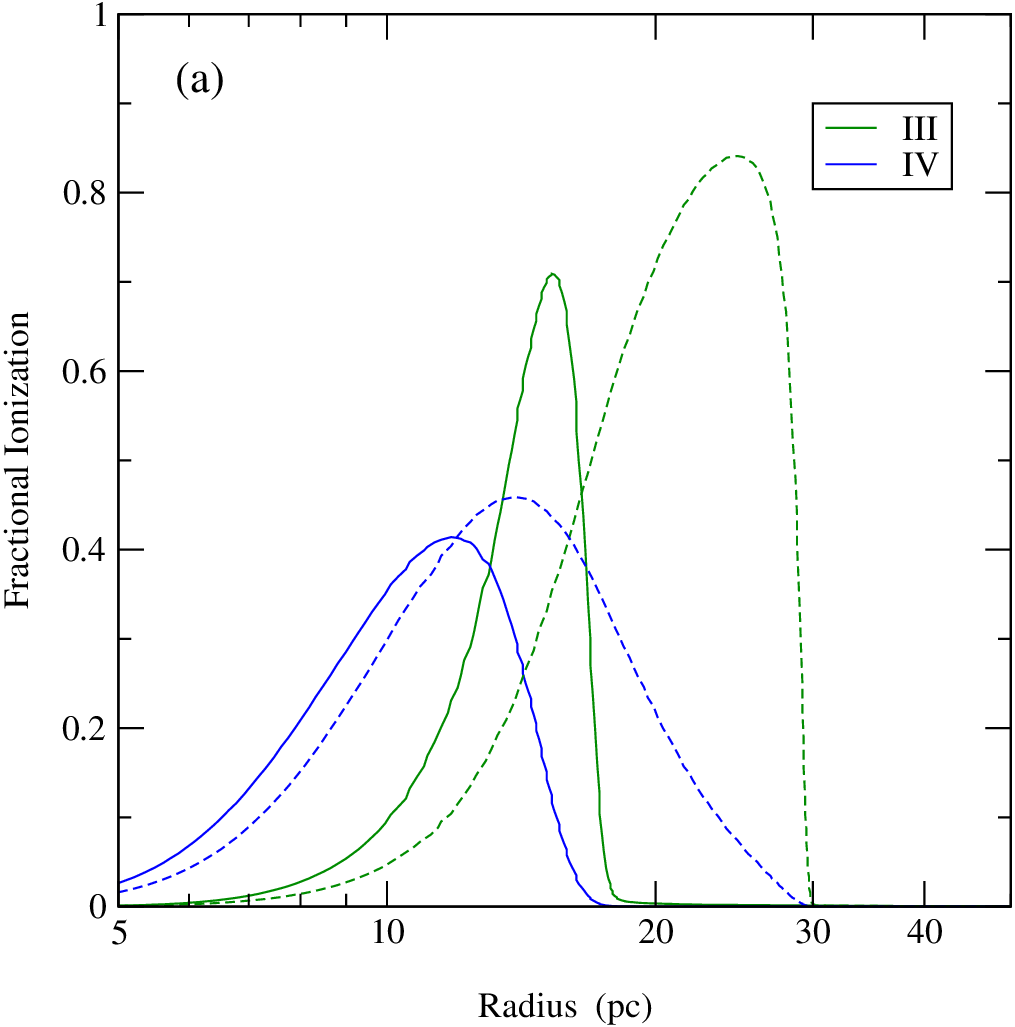}{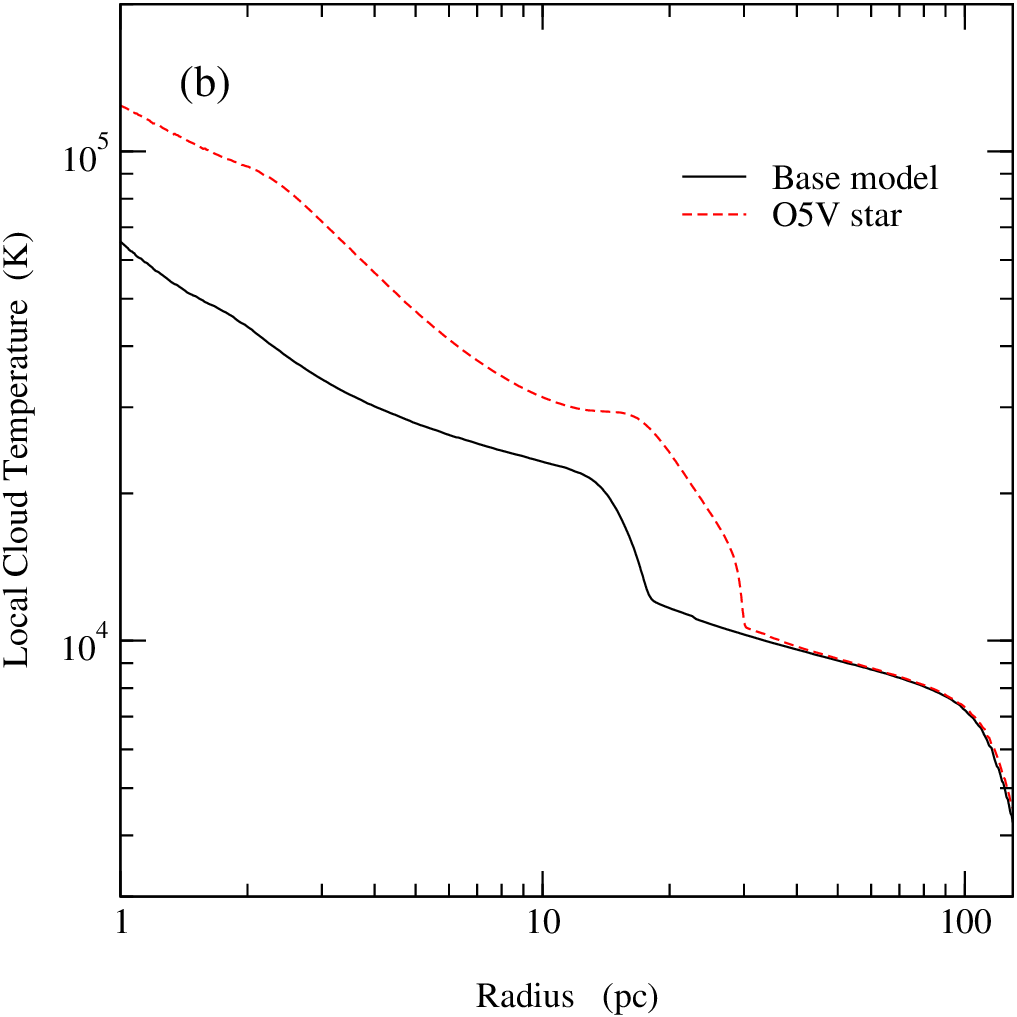}
\caption{
Comparison between the base model, with the B2Ib star (continuous line) 
and the model with the O5V star (dashed line). 
a) The oxygen ionic structure in the cloud (O$^{2+}$ and O$^{3+}$).
b) Local temperature in the cloud.
} \label{oxygen}
\end{figure}

{\bf Varying the Stellar Companion (the O5V star):}  To test the effect of the stellar companion on the IR lines 
we ran the base model with an O5V star replacing the B2Ib supergiant. The temperature and luminosity of the O5V star, 
42000~K and 3.2$\times$10$^{39}$ erg~s$^{-1}$, were chosen to match the model used by KWZ. 
The results (the lower-right panel of Fig.~\ref{predictions}) show that the O5V star 
has a very strong impact on the high ionization lines. The luminosity of the [O~IV] line increases by a factor of 3, 
and the [Ne~V] line by factor of 2. If the companion is in fact an O5V star, 
our measured Optical Monitor (OM) and {\it GALEX} fluxes from Paper~1 are under-luminous 
and are likely affected by reddening. Inclusion of an O5V star also has a very different effect on the [O~IV] 
versus the He~II line. Indeed, the He~II luminosity is predicted to increase by only 11\%. 
The strong impact of the O5V star can be explained if the O star produces copious O$^{++}$ ions (i.p. 35~eV), 
which are in turn ionized further by the ULX. The O star thus acts as a catalyst to the higher-energy photons from the ULX.  
A comparison between the oxygen ions structure for these models is presented in Figure~\ref{oxygen}a. 
The number of photons in the 24.6$-$54.4~eV interval is three times higher for the O5V star 
(7.27$\times$10$^{48}$~s$^{-1}$ compared to 2.42$\times$10$^{48}$~s$^{-1}$) than the B2Ib supergiant.  
Figure~\ref{oxygen}b shows that the local cloud temperature for the O5V star is also about two times higher 
in the region where the O$^{3+}$ ions are produced (at offsets of 10$-$20~pc). 
Indeed, the [O~III] $\lambda$5007 luminosity estimated from CLOUDY increases by a factor of six 
when the O5V star is used instead of the B2Ib supergiant. 

The companion star therefore has an indirect but strong impact on the [O~IV] line emission.  
A similar effect can be seen on a smaller scale for the [Ne~V] line,
which has an even higher ionization potential (97~eV). Thus, the [O~IV] line does not act simply 
as a high energy photon counter as is the case with the He~II line, but can also provide information 
about abundances of lower energy photons. This effect can be used to explain the detection of [O~IV] 
in star-forming regions. In this scenario, [O~IV] can be produced by relatively faint sources of soft X-rays 
and UV photons hidden in the star forming region, including such sources as ULXs, X-ray binaries, or SNR.  
This is a significant result, since the origin of [O~IV] in star forming galaxies has been the subject of debate 
in the literature \citep{lutz98,scha99}.

Because our CLOUDY modeling overpredicts most of the high ionization line fluxes when an OV5 star is used, 
we conclude that an OV5 star is likely not the companion star for the Holmberg~II ULX.   
On the other hand, later-type stars, such as an O8 star (or later), 
might better predict the [O~IV] flux in the base model.   
In support of this, if the ULX is powered by a stellar-mass BH, 
the companion required to reproduce the observed X-ray luminosities must be very massive \citep{pod03}.   
Thus a stellar mass black hole could explain both the [O~IV] luminosity and the very high X-ray luminosity.  
The fact that the Holmberg II ULX resides just outside a UV-bright region of star formation (see Paper~1), 
also suggests that the environment is well suited for formation of such massive stars 
\citep[similar to NGC4559 X7, see][]{sor07}. However, without the black hole mass constrained, 
an IMBH with a massive companion is still also a plausible option.

\subsection{Varying the Geometry}

As mentioned at the beginning of this section, KWZ found evidence from low-ionization optical lines 
that the ionized nebula is matter-bounded, at least in the east. This is also suggested by the IR line maps 
presented in Paper~1 and by the over-prediction of the low-ionization lines by the radiation bounded base SED model above. 
Paper~1 also shows that the low ionization line fluxes are likely contaminated by circum-stellar emission.
We also noticed that the absorption suggested by the X-ray spectrum is too low compared 
to that implied by a radiation-bounded geometry. More precisely, for our base SED model, 
the simulation requires a column density of 4.2~$\times$10$^{21}$ cm$^{-2}$, 
which is more than ten times larger than the measured values from spectral fits (see also the beginning of this section).  
This suggests that the nebula is also matter-bounded in the line-of-sight direction.

\begin{figure}
\epsscale{0.9}
\plottwo{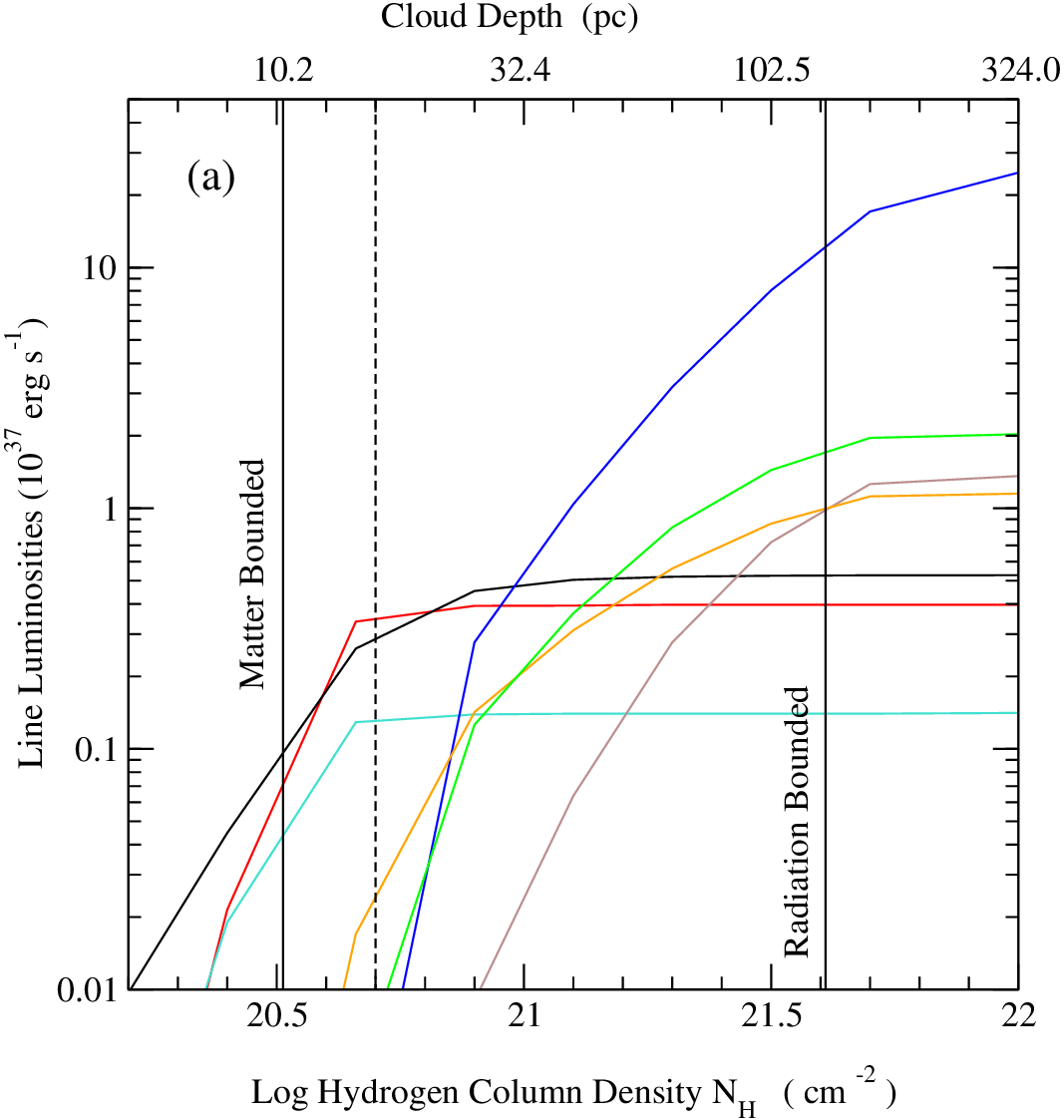}{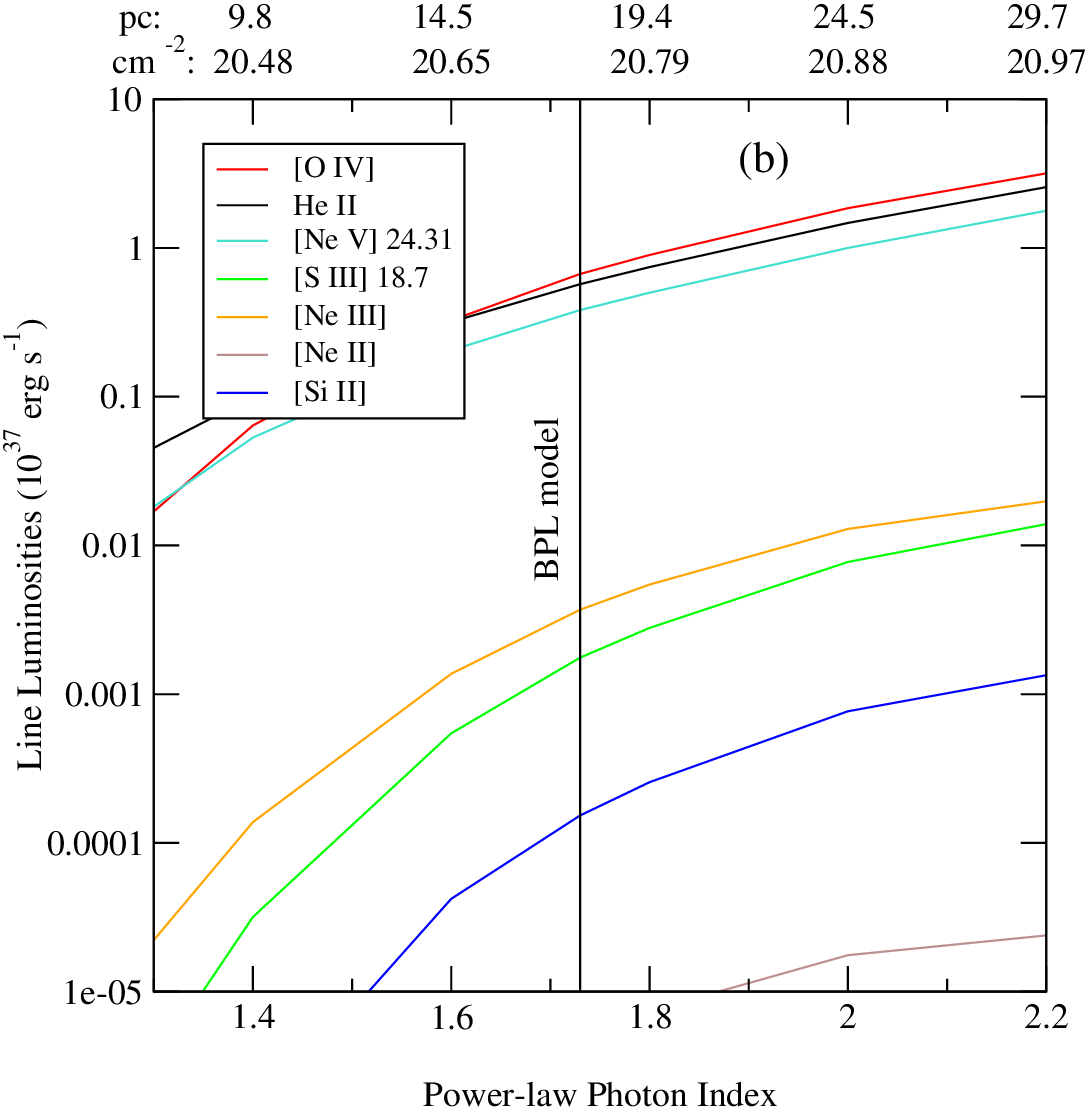}
\caption{
This figure illustrates the difference between a matter bounded and radiation bounded nebula.
The lines are only labeled in the right panel.
a) The dependence of the simulated lines from CLOUDY on the cloud hydrogen column density for the base model.
The equivalent depth of the spherical cloud is also marked on the top X-axis. 
Both vertical continuous lines mark the column density for the base model,
but one value was obtained by fitting the X-ray data with the base model (matter bounded nebula),
while the other is the column in the CLOUDY simulation where the local temperature reached 4000~K (radiation bounded).
We also marked with a vertical dashed line the limit where the predictions for
the high excitation lines are still compatible with observations.
b) The simulated lines using the family of broken power-law models fitted to the X-ray data.
The best fit broken PL model (BPL) is marked by the vertical line.
The bottom X-axis shows the low-energy photon index for the models. 
On the top X-axis we mark the corresponding log column densities (lower values) and cloud thickness (upper values).
All these models are matter-bounded. 
} \label{matter}
\end{figure}

To simulate a matter-bounded geometry, we first ran the base SED model with the geometry 
described by a spherical cloud of varying thickness, so that the total hydrogen column 
ranges between 10$^{20}$ and 10$^{22}$~cm$^{-2}$. These results are plotted in Figure~\ref{matter}a.   
The column obtained from the X-ray fit is shown as a vertical line and labeled ``Matter Bounded''.
For this value, the predicted fluxes for both [O~IV] and He~II lines are much lower than the observed values. 
Therefore, this simple matter bounded geometry using the base SED model is not sufficient to reproduce the observed lines.  
As Figure~\ref{matter}a shows, the minimum column density required for the base model 
to predict the measured values for the high excitation lines is $\sim$5~$\times$10$^{20}$, 
or a cloud thickness of 16~pc (the dashed vertical line). This values is within the 15-25~pc radii 
observed in the He~II emission line maps from KWZ. Therefore the matter bounded case could work, 
if the observed X-ray column density were slightly higher than the X-ray predicted value of 3.26~$\times$10$^{20}$~cm$^{-2}$.

We next replaced the Modified PLMCD model for the BPL model as the X-ray fit for the SED.  
The broken power-law model is better suited than the Modified PLMCD model to test the matter bounded geometry. 
By varying this absorption parameter in the model 
within the errors allowed by the fit, we obtained a family of broken-power models, 
all consistent with the X-ray data, but predicting very different slopes 
in the ionizing part of the SED (the photon index varies between 1.3 and 2.2).  
A subset of the broken power-law models are shown in Figure~\ref{sed}. 
We re-ran our CLOUDY simulation for these models with the corresponding column densities from the X-ray fits. 
The results are plotted in Figure~\ref{matter}b. The best fit broken power-law X-ray model (BPL) 
is shown as a vertical line in this figure. The predicted line luminosities for this model 
are given in the lower-left panel of Figure~\ref{predictions}. 

As was the case with the original, 
radiation bounded base SED model, the predictions are very good for the high-ionization lines. 
The cloud radius is 17.7~pc, which is consistent with the size of the nebula found by KWZ in the east and south.  
However, contrary to the original radiation-bounded base model, the ULX makes no contribution to the low-ionization lines. 
This is true even for the largest column density, 9.2$\times$10$^{20}$ cm$^{-2}$  
(a 29.7~pc cloud radius for a gas density of 10~cm$^{-3}$). However, the spectral maps in Figure~3 of Paper~1 
show that the ULX is likely to contribute at least to [Ne~III] and probably [Si~II].

Thus we conclude that the matter-bounded geometry alone does not fit all of the mid-IR spectral observations well.  
In fact, these results suggest that the geometry is complicated, being asymmetric 
and including both radiation and matter bounded geometries.  KWZ found similar results.  
They find that to the south and east, He~II and H$\beta$ emission suggests a matter-bounded geometry: 
the low excitation lines are absent here but detected in the west at larger offsets. 
Our CUBISM maps from Paper~1 suggest something similar, though the spatial resolution here is limited 
compared with the {\it HST} observation. The fact that the low-ionization line luminosities 
predicted by the matter-bounded models are much lower than the ones predicted by the radiation-bounded geometry 
suggests that a combination of the two geometries would yield predicted values that are very close to the observations.

\section{Shock Models}

The radiation-bounded/matter-bounded geometric scenario presented in Section~3 to explain the mid-IR line fluxes 
in the Holmberg~II ULX presumes photoionization as the source of the line emission.  
However, shocks could also produce such lines. Indeed, a combination of X-ray photoionization and shocks 
was recently proposed to explain the morphology of optical lines observed in the vicinity of LMC~X-1 \citep{cooke07}.  
In this case the shocks are likely driven by a jet. \citet{abol07} performed optical spectroscopy on eight ULXs 
with nebular counterparts, including the Holmberg~II ULX, and found evidence for shocks in all cases 
based on the [S~II] $\lambda\lambda$6717,6731/H$\alpha$ ratio. However, 
using high-ionization lines such as He~II and [O~III], these authors also found that Holmberg~II ULX 
and three other ULXs required photoionization to sufficiently explain the optical ratios.

\citet{leh05} found evidence of radial velocity variations of $\pm$~50~km~s$^{-1}$ 
from a region similar of size to the He~III nebula. It is therefore possible 
that part of the ionization is caused by shocks. \citet{lutz98} suggested that velocity shocks 
of 100$-$200~km~s$^{-1}$ can produce significant [O~IV] emission. However, when we also take into account the optical data, 
and in particular the detection  of the He~II recombination line, our data provide strong evidence 
{\it against} shocks as the source of the [O~IV] emission. This is primarily because the He~II 
is not easily produced by shocks. Instead, the simultaneous detection of these lines 
argues for the presence of significant radiation over the He~II edge \citep[see also ][]{scha99}, 
and therefore suggests photoionization as the most likely line production mechanism.

\begin{figure}
\epsscale{1.0}
\plottwo{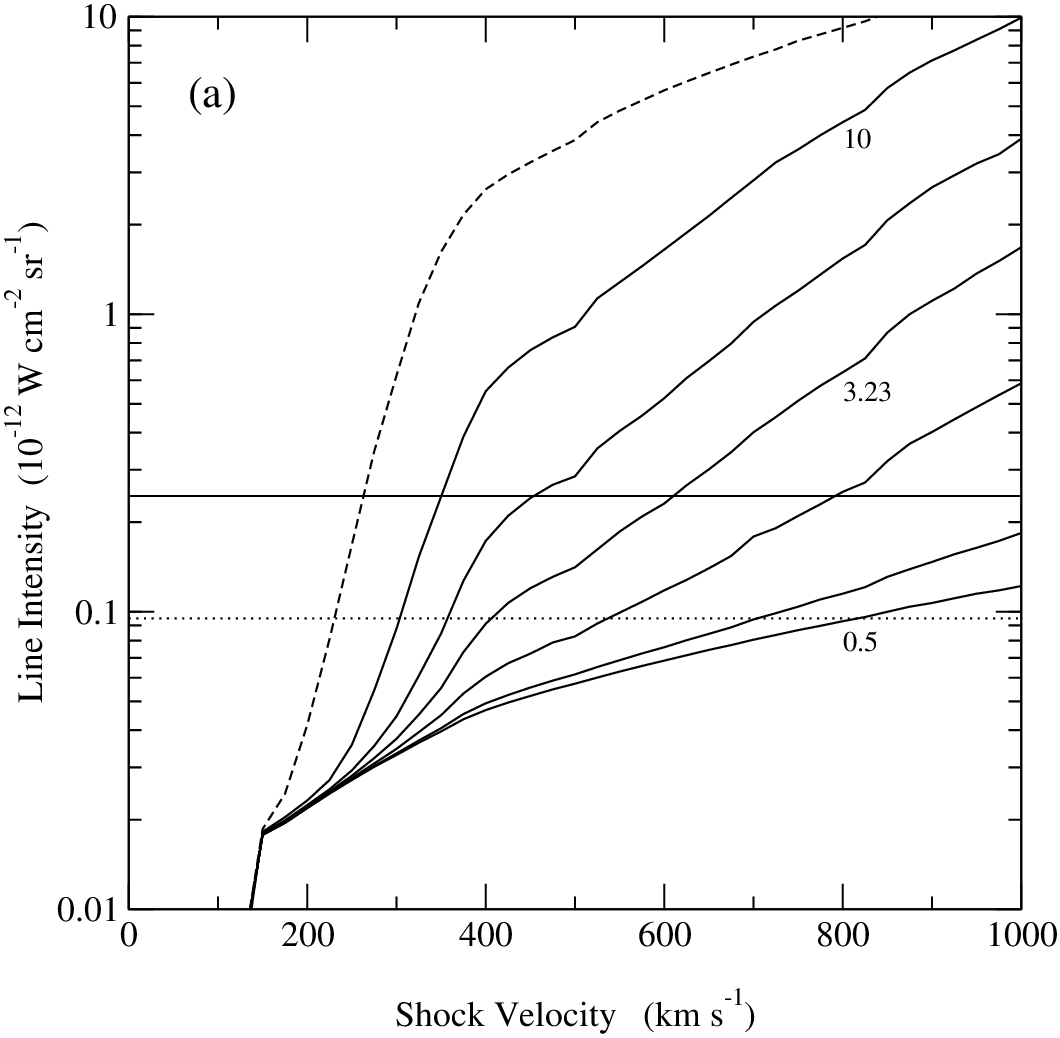}{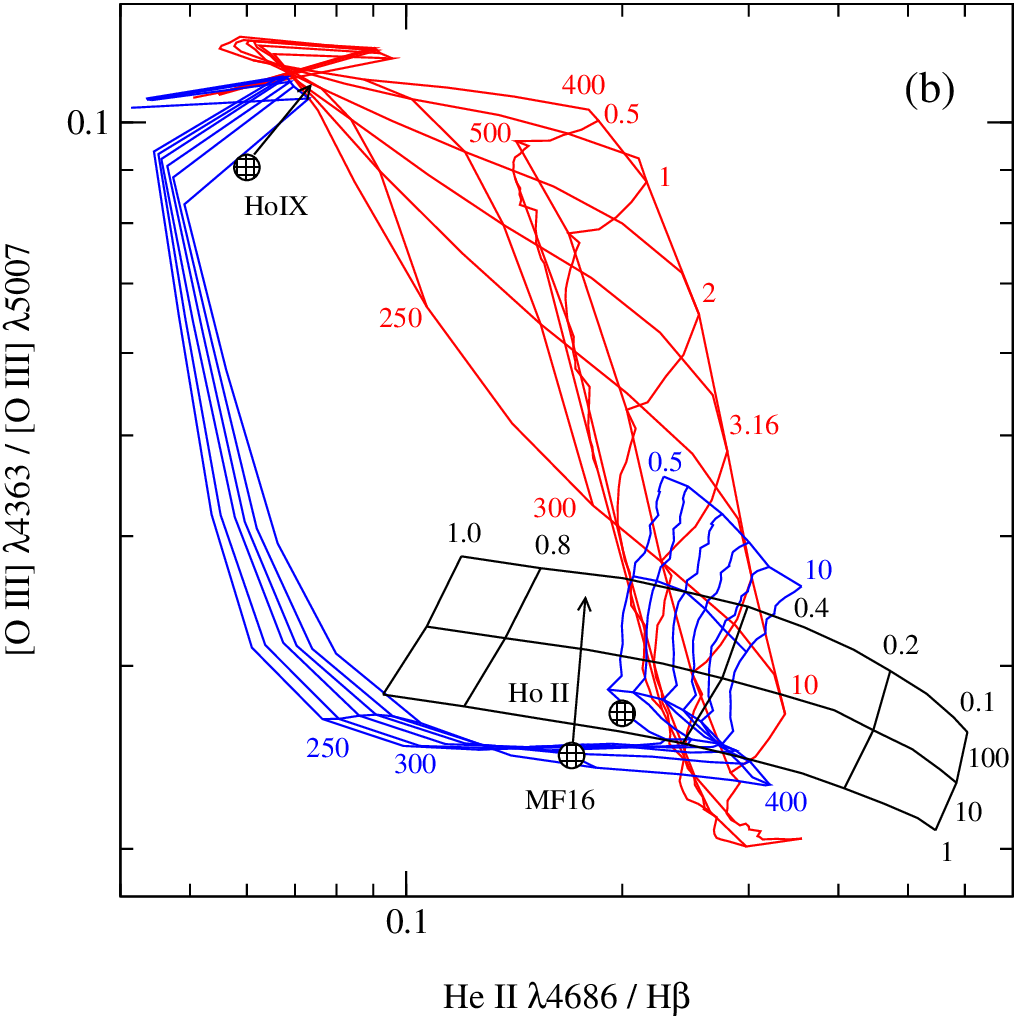}
\caption{
Mappings~III shocks and photoionization models.
The shock grids are plotted for magnetic parameters: 0.5, 1, 2, 3.23, 5 and 10 $\mu$G~cm$^{3/2}$ 
and for shock velocities 125, 250, 300, 400, 500, 700 and 1000~km~s$^{-1}$.
a) [O~IV] line intensity shocks predictions. 
The shocks plus precursor models vary little with the magnetic parameter, 
and we only show the model with a nominal equipartition value (3.23~$\mu$G~cm$^{3/2}$, the dashed line).
The horizontal lines correspond to the brightest (continuous) and respectively faintest (dotted) 
pixels in the [O~IV] line map in Figure~3 of Paper~1.
b) Optical lines diagnostic with shocks and photoionization models.
The pure shock models are in red, shocks plus precursor in blue and the photoionization grid in black.
For the photoionization grid we varied the density (1, 10 and 100 cm$^{-3}$),
and the inner disk temperature (0.1, 0.2, 0.4, 0.8 and  1.0~keV), and assumed a metallicity of 0.1~Z$_\odot$.
For Holmberg~IX ULX and MF16 we show with arrows the approximate locations for Solar metallicity.
While Holmberg~IX ULX is consistent with shocks with modest velocities ($<$~100~km~s$^{-1}$),
Holmberg~II ULX and MF16 seem to require very high velocities. They are however consistent
with photoionization. 
} \label{shocks}
\end{figure}

Nevertheless, we checked to see if shock models could explain the mid-IR line luminosities and ratios 
observed in Holmberg~II. We used Mappings~III grids from \citet{allen08} to construct diagnostic diagrams.  
The predicted [O~IV] emission from shocks is shown in Figure~\ref{shocks}a. We show both pure shocks 
and shocks with ionized precursors at a gas density of 1~cm$^{-3}$, and metallicity 0.1~Z$_\odot$.  
For these models, the parameters varied are the shock velocity and the ``magnetic parameter'' B/n$^{1/2}$, 
where B is the magnetic field and n is the gas density \citep{dop96}. 

The CUBISM measured fluxes from the brightest and faintest pixels in the [O~IV] map 
(yellow and blue pixels in Fig.~3 of Paper~1, respectively) are shown as horizontal lines in Figure~\ref{shocks}a. 
We note that the maximum intensity of the line is likely higher than these levels. 
Indeed the [O~IV] emission is not spatially resolved 
and the plotted levels are actually averages over the size of the LH pixel (67.5~pc), 
rather than the maximum from each of the averaged slits. In spite of this, 
the levels plotted in Figure~\ref{shocks}a requires significant shock velocities 
and such velocities ($>$~200~km~s$^{-1}$) were not detected by \citet{leh05}.

For optical lines we chose the He~II and the [O~III] $\lambda\lambda$4363,5007 lines, 
as per \citet{evans99}. The ratio [O~III] $\lambda$4363 / [O~III] $\lambda$5007 
is sensitive to the gas temperature. Using the data from \citet{leh05} 
we calculated the ratio ([O~III] $\lambda$4959 $+$ [O~III] $\lambda$5007) / [O~III]~$\lambda$4363~$=$~50, 
and estimate a temperature of 17800~K \citep{ost06}. This is consistent 
with the temperature of the gas in the region of the cloud where O$^{++}$ ions are produced (see Fig.~\ref{oxygen}b).  
For the He~II/H$\beta$ ratio we chose an average (0.2) between the measurements of KWZ, \citet{pak02}, and \citet{leh05}. 
The shock grids for the optical lines are plotted in Figure~\ref{shocks}b. 
For comparison, we also plot a photoionization grid obtained with Mappings~III, 
using parameters similar to those used in the CLOUDY modeling.
Finally, we plot previously published data for two other ULXs with detected nebulae, 
Holmberg~IX ULX and MF16 in NGC~6946 \citep{abol07}. While the Holmberg~IX ULX data 
is consistent with shocks with modest velocities ($<$~100~km~s$^{-1}$), 
Holmberg~II ULX and MF16 seem to require very high velocities. Even for the shock plus precursor models, 
velocities in excess of 300~km~s$^{-1}$ are required to reproduce the observed line ratios
and \citet{leh05} find velocities of only 50~km~s$^{-1}$.

In summary, some contribution from shocks to the high excitation lines detected from Holmberg~II ULX 
cannot be ruled out completely. However, if there is some contribution from shocks 
to the high ionization emission line luminosities, it is likely very small, 
since both He~II and [O~IV] lines require shocks with velocities in excess of 200~km~s$^{-1}$,
and such velocities were not seen by \citet{leh05}.

\section{Discussion of the Modeling Results}

We find that the {\it Spitzer} observations of the Holmberg~II ULX and especially the detection of the [O~IV] line 
is consistent with photoionization by radiation from the ULX. 
Of all the lines examined, only the [O~IV] was shown in Paper~1 to be correlated with the ULX.  
However, we also found in Paper~1 that some contribution by the ULX to the [Si~II], [S~III], 
and [Ne~III] emission line flux is also likely. The modeling results presented here are based 
on the observations presented in Paper~1 and are summarized as follows.

\begin{enumerate}

\item {\bf The Base Model (Modified PLMCD):} The CLOUDY simulations show that the {\it Spitzer} spectrum 
is consistent with photoionization by the accretion powered emission from the ULX.  
We find that limited sensitivity might explain the relatively low fluxes observed 
for some low-ionization lines. The base SED model predicts a slightly lower [O~IV] luminosity than the measured values 
for the radiation bounded case (Fig.~\ref{predictions}), and even lower for the matter-bounded case (Figure~\ref{matter}a). 
Previous X-ray results have shown that the accretion disk temperature is likely lower 
than in our base model (KWZ obtained $\sim$0.2~keV). Such a disk temperature 
will increase the predicted line luminosity to match our measurements (Fig.~\ref{param1}a). 
A slightly higher metallicity (0.2~Z$_\odot$) can yield better fits to both the [O~IV] and He~II data 
for this base model. This is because the He~II line is already over-predicted and raising the metallicity 
will not increase this discrepancy (Fig.~\ref{param2}a). 
In support of this, a metallicity of 0.17 was measured by \citet{pil04}. We note that while this model 
reproduces the spectra very well, it requires absorbing columns in the X-rays 
that are inconsistent with the observations.

\item {\bf The PLMCD Model:} We find that the widely used PLMCD model, when extrapolated to the UV, 
is {\it not} consistent with the observations. The high ionization lines predicted using this model 
are at least ten times larger than observed. In this case lowering the metallicity to match the [O~IV] 
will not work, because He~II will still be much over-predicted. This model has been used extensively 
to fit the X-ray data from ULXs \citep[e.g. ``cool disks''][]{mill04}. 
However, the X-ray community has more recently begun to fit these spectra 
with more physically motivated models \citet[e.g.]{free06, done06, kun07, sor07, sor08}, 
which appear to be consistent with  high quality ULX X-ray data \citep{ber08}.  
That the PLMCD model is inconsistent with the {\it Spitzer} observations of the Holmberg~II ULX
supports these previous X-ray studies.

\item {\bf Matter vs. Radiation Bounded Geometry:}  The intensity and morphology of the IR lines 
provides spatial information about the emission line nebula around the Holmberg~II ULX.  
The geometry consistent with our infrared data (and previously published optical data) 
is approximately spherical but not symmetrical. The CUBISM maps show that to the east and south the nebula 
shows only high excitation lines that are likely matter-bounded, while in the west, 
the detection of lower excitation lines might suggest a radiation-bounded cloud.  
However, observational effects, such as aperture and sensitivity effects, preclude any definitive conclusion 
on this model. The discrepancies found in Section~3.1 between the CLOUDY model
and the measured values for the low-excitation lines suggest that the matter-bounded geometry
is at play at least in part (probably to east and south as found from the spectral maps).
Moreover, the absorption measurements from the X-ray data require the cloud  
to be at least matter-bounded toward our line of sight. Therefore, we find that 
both matter bounded and radiation bounded geometries are likely at work 
to describe the observed spectra in both the Mid-IR and X-rays.  

\item {\bf Shock Models:} The shocks models for both IR and optical emission lines 
show that shocks with velocity $>$200~km~s$^{-1}$ are required to generate 
both the He~II and [O~IV] lines observed in Holmberg~II. The detected radial velocity variations 
around the ULX are of the order of $\pm$50~km~s$^{-1}$ \citep{leh05}. Contribution from shocks to these lines 
is therefore likely to be small.

\item {\bf True X-ray Luminosity:}  The ULX luminosity estimated using photoionization modeling 
is independent of the estimate based on the X-ray flux, which can be affected by both absorption 
and beaming \citep{king01}. The nebular emission line luminosities, on the other hand, 
are time-averaged true luminosities and can provide information about the geometry of emission 
and the surrounding ISM.

It is now widely accepted that ULXs are variable on timescales from days to years.  
These timescales are much shorter than the ionization equilibrium timescales in the nebula.  
\citet{chiang96} investigated time-dependent photoionization of H and He for supersoft X-ray sources. 
The He~II ionization equilibrium at the 90\% level is reached in $\sim$0.7~$\tau$$_R$ after source turn-on, 
where $\tau$$_R$~$=$~3000~yr is the recombination time for He$^{++}$ from KWZ. For periodic sources, 
it was found that the He~II line luminosity decreases significantly with increasing source period. 
For example, for a period of 10~$\tau$$_R$ and a duty cycle of 10\%, the estimated luminosity 
is 4.7\% of that expected from a steady source with the same peak luminosity.  
That is a 53\% deviation from a linear scaling law (10\% of the steady state). 
This implies that if the Holmberg~II ULX is variable, the estimates of KWZ based on the He~II line 
are actually lower limits of the time-averaged luminosity. 

For O$^{+3}$, the recombination time is shorter, 
$\tau$$_R$~$\approx$~100~yr \citep{ost06}, for the same ionization conditions: 
electron density 10~cm$^{-3}$ and temperature 20000~K. We expect therefore that the [O~IV] line 
reaches equilibrium in a shorter time ($<$~100~yr), but still long enough to provide 
a time-averaged estimate of the true luminosity of the ULX over $\sim$10~$-$~100 yr. 
We note that light travel time (53~yr for a O$^{+3}$ nebula of 16~pc), is an important factor in this case,
because it is comparable to the recombination time.  

Finally, the morphology of the emission lines, from both the previously published optical data 
and our IR data suggest emission that is nearly isotropic, being inconsistent with narrow beaming. 
Moreover, photoionization modeling with our base model is consistent with the detected high ionization lines,
the true bolometric luminosity of this model is $>$10$^{40}$ erg~s$^{-1}$. This is well above the Eddington
luminosity for a stellar-mass BH. In addition, some observational results indicate that the BH in Holmberg~II 
is likely much larger than a stellar mass black hole (see the conclusion below).

\item {\bf Impact on IR Studies of Starbursts:}  We find that the predicted [O~IV] line luminosity 
is significantly affected by the type of the companion star. Faint [O~IV] has been detected 
in many starbursts and star-forming regions, but its origin is still under debate. 
Our analysis provides a new mechanism for this emission, suggesting that the [O~IV] line 
could be produced by relatively faint X-ray sources including X-ray binaries, SNR or ULXs in starburst galaxies.

\end{enumerate}

\section{Conclusion} 

The bolometric luminosity of the base model 
is 1.33~$\times$10$^{40}$ erg~s$^{-1}$. For accretion within the Eddington limit, 
the mass of the central BH is at least 85~M$_\odot$. The estimate based on the inner disk temperature 
(0.38~keV) in the MCD model gives a much larger mass, 994~M$_\odot$. Finally, scaling from AGN estimates, 
the [O~IV] emission predicts a mass as large as 10$^4$~M$_\odot$.   
All of the estimates presented here presume the black hole is accreting within the Eddington limit.  
However, while our analysis excludes strong beaming in Holmberg~II ULX, 
a sub-Eddington IMBH or a super-Eddington stellar-mass BH are both plausible ionization mechanisms 
that stem from our analysis. Indeed we show in Section 3.3 that a massive companion (e.g. B2Ib star) 
and a stellar mass black hole can reproduce the [OIV] luminosities see in the Spitzer spectra, 
as can an IMBH and a similar companion. The CLOUDY modeling results alone cannot discriminate between these two sources 
or the black hole mass estimates, since we have no way of constraining the accretion rate.

\acknowledgments

C. T. B. thanks Richard Mushotzky and Lisa Winter for helpful discussions, 
and Nicholas Sterling and Marcio Mel{\'e}ndez for their help with CLOUDY. 
R. P. D. gratefully acknowledges financial support from the NASA Graduate Student Research Program.
This work is based on observations made with the Spitzer Space Telescope, 
which is operated by the Jet Propulsion Laboratory, California Institute of Technology under a contract with NASA.
We thank the referee for very helpful and constructive comments that have significantly improved this paper.



\begin{deluxetable}{l|cccccc}
\tablecolumns{6}  
\tablewidth{0pt} 
\tabletypesize{\tiny}
\setlength{\tabcolsep}{0.05in}
\tablenum{1} 
\tablecaption{X-ray model fits\label{table2}}   
\tablehead{
\colhead{Model} & \colhead{N$_H$} & \colhead{kT$_{in}$/$\Gamma$$_1$} & \colhead{$\Gamma$/$\Gamma$$_2$} & \colhead{$\Delta\chi^2$/dof} & \colhead{L} & \colhead{MCD flux} \\
\colhead{(1)} & \colhead{(2)} & \colhead{(3)} & \colhead{(4)} & \colhead{(5)} & \colhead{(6)} & \colhead{(7)} \\
\colhead{} & \colhead{(10$^{20}$ cm$^{-3}$)} & \colhead{(keV)} & \colhead{} & \colhead{} & \colhead{(10$^{40}$ erg s$^{-1}$)} \\
}
\startdata

Modified PLMCD	   &  3.26$\pm$0.56  &	0.38$\pm$0.02  &  2.54$\pm$0.02	 &  1.02/135, 0.974/713	&  1.11 &  0.49 \\
PLMCD	           &  10.4$\pm$0.70  &	0.27$\pm$0.02  &  2.42$\pm$0.05	 &  1.1386/849	        &  2.28 &  0.12 \\
BPL                &  5.49$\pm$0.78  &	1.73$\pm$0.10  &  2.57$\pm$0.02	 &  1.0316/849	        &  1.41 &  0.60 \\

\enddata

\tablecomments{  
(1): X-ray model, as described in Section~2.
(2): Intrinsic hydrogen column density. 
The Galactic column density from KWZ (3.42$\times$10$^{20}$ cm$^{-2}$) was added separately. 
(3): Inner disk temperature for the MCD component. For the broken power-law model (BPL), this is 
the first (low energies) photon index parameter. The break is at 1.0~keV.
(4): Photon index for the power-law component. 
For the broken power-law model this is the second (high energies) photon index parameter.
(5): Reduced $\chi^2$ values for the fit and the number of degrees of freedom. 
For the Modified PLMCD model, we quote two values for each separately fitted component (see text for details).
(6): Unabsorbed (intrinsic) luminosities between 0.1 and 10 keV.
(7): MCD component unabsorbed flux as fraction of the total flux for the PLMCD model.
For the other two models, this simply the ratio of the flux between 0.1 and 1.0~keV
to the flux in the whole range (0.1$-$10~keV)
}
\end{deluxetable}


\begin{deluxetable}{c|ccc|ccc}
\tablecolumns{6}  
\tablewidth{0pt} 
\tabletypesize{\tiny}
\setlength{\tabcolsep}{0.05in}
\tablenum{2} 
\tablecaption{Measured infrared lines\label{table1}}   
\tablehead{
\colhead{} & \multicolumn{3}{c}{Standard Aperture} & \multicolumn{3}{c}{Small Aperture} \\
\tableline
\colhead{Line} & \colhead{Flux} & \colhead{S/N Ratio} & \colhead{L} & \colhead{Flux} & \colhead{S/N Ratio} & \colhead{L} \\
\colhead{(1)} & \colhead{(2)} & \colhead{(3)} & \colhead{(4)} & \colhead{(5)} & \colhead{(6)} & \colhead{(7)} \\
}
\startdata

$[$Ne II$]$  12.81~$\mu$m  &  $<$6.08	     &  ...   &  $<$0.68        &  $<$1.98	  &	...  &	$<$0.22	        \\
$[$Ne III$]$ 15.56~$\mu$m  &  12.15$\pm$3.4  &  4.60  &  1.35$\pm$0.39  &  5.46$\pm$1.37  &	7.6  &	0.61$\pm$0.15	\\
$[$S III$]$  18.71~$\mu$m  &  2.64$\pm$1.14  &  4.64  &  0.29$\pm$0.13  &  3.02$\pm$1.16  &	4.7  &	0.34$\pm$0.13	\\
$[$Ne V$]$   24.32~$\mu$m  &  $<$3.88	     &  ...   &  $<$0.43        &  $<$1.16	  &	...  &	$<$0.13	        \\
$[$O IV$]$   25.89~$\mu$m  &  7.01$\pm$1.60  &  10.16 &  0.78$\pm$0.2   &  3.63$\pm$0.91  &	44.4 &	0.40$\pm$0.1	\\
$[$S III$]$  33.48~$\mu$m  &  $<$4.39	     &  ...   &  $<$0.49        &  $<$1.53	  &	...  &	$<$0.17	        \\
$[$Si II$]$  34.82~$\mu$m  &  21.91$\pm$5.48 &  9.90  &  2.44$\pm$0.61  &  5.87$\pm$1.47  &	7.2  &	0.65$\pm$0.16	\\

\enddata

\tablecomments{   
We show line fluxes for the standard (slightly smaller than the SH map) 
and the small aperture (4 LH pixels) as defined in Section~2 of Paper~1. 
Fluxes are in 10$^{-22}$~W~cm$^{-2}$, luminosities (L) in 10$^{37}$~erg~s$^{-1}$.
If the measurements errors are smaller than the absolute calibration accuracy of 25\%, 
the latter were used. For nondetections we show 3$\sigma$ upper limits.
}
\end{deluxetable}


\begin{thebibliography}{}

\bibitem[Abolmasov et al.(2007)]{abol07} Abolmasov, P., Fabrika, S., Sholukhova, O., \& Afanasiev, V.\ 2007, Astrophysical Bulletin, 62, 36 
\bibitem[Allen et al.(2008)]{allen08} Allen, M.~G., Groves, B.~A., Dopita, M.~A., Sutherland, R.~S., \& Kewley, L.~J.\ 2008, \apjs, 178, 20 
\bibitem[Berghea et al.(2008)]{ber08} Berghea, C.~T., Weaver, K.~A., Colbert, E.~J.~M., \& Roberts, T.~P.\ 2008, \apj, 687, 471 
\bibitem[Berghea et al.(2009)]{ber09} Berghea, C.~T., Dudik, R.~P., Weaver, K.~A., \& Kallman, T.~R.\ 2009, arXiv:0904.1837 
\bibitem[Chiang \& Rappaport(1996)]{chiang96} Chiang, E., \& Rappaport, S.\ 1996, \apj, 469, 255 
\bibitem[Cooke et al.(2007)]{cooke07} Cooke, R., Kuncic, Z., Sharp, R., \& Bland-Hawthorn, J.\ 2007, \apjl, 667, L163 
\bibitem[Dale et al.(2006)]{dale06} Dale, D.~A., et al.\ 2006, \apj, 646, 161 
\bibitem[Dopita \& Sutherland(1996)]{dop96} Dopita, M.~A., \& Sutherland, R.~S.\ 1996, \apjs, 102, 161 
\bibitem[Done \& Kubota(2006)]{done06} Done, C., \& Kubota, A. 2006, \mnras, 371, 1216
\bibitem[Evans et al.(1999)]{evans99} Evans, I., Koratkar, A., Allen, M., Dopita, M., \& Tsvetanov, Z.\ 1999, \apj, 521, 531 
\bibitem[Fabbiano(1989)]{fabbiano} Fabbiano, G.\ 1989, \araa, 27, 87 
\bibitem[Freeland et al.(2006)]{free06} Freeland, M., Kuncic, Z., Soria, R., \& Bicknell, G.~V.\ 2006, \mnras, 372, 630 
\bibitem[Genzel et al.(1998)]{gen98} Genzel et al. 1998, \apj, 498, 579
\bibitem[Goad et al.(2006)]{goad06} Goad, M.~R., Roberts, T.~P., Reeves, J.~N., \& Uttley, P.\ 2006, \mnras, 365, 191 
\bibitem[Kaaret et al.(2003)]{kaa03} Kaaret, P., Corbel, S., Prestwich, A.~H., \& Zezas, A.\ 2003, Science, 299, 365 
\bibitem[Kaaret et al.(2004)]{kaa04} Kaaret, P., Ward, M., \& Zezas, A. 2004, \mnras, 351, L83
\bibitem[Kaaret \& Corbel(2009)]{kaa09} Kaaret, P., \& Corbel, S.\ 2009, \apj, 697, 950
\bibitem[King et al.(2001)]{king01} King, A.~R., Davies, M.~B., Ward, M.~J., Fabbiano, G., \& Elvis, M. 2001, \apj, 552, L109
\bibitem[King(2008)]{king08} King, A.~R.\ 2008, \mnras, 385, L113 
\bibitem[Kuncic et al.(2007)]{kun07} Kuncic, Z., Soria, R., Hung, C.~K., Freeland, M.~C., \& Bicknell, G.~V.\ 2007, IAU Symposium, 238, 247 
\bibitem[Lehmann et al.(2005)]{leh05} Lehmann, I. et al. 2005, A\&A, 431, 847
\bibitem[Lutz et al.(1998)]{lutz98} Lutz, D., Kunze, D., Spoon, H.~W.~W., \& Thornley, M.~D. 1998, A\&A, 333, L75
\bibitem[Miller et al.(2004)]{mill04} Miller, J.~M., Fabian, A.~C., \& Miller, M.~C. 2004, \apj, 614, L117
\bibitem[Miller \& Colbert(2004)]{millco} Miller, M.~C., \& Colbert, E.~J.~M.\ 2004, International Journal of Modern Physics D, 13, 1 
\bibitem[Miller et al.(2005)]{mill05} Miller, N.~A., Mushotzky, R.~F., \& Neff, S.~G. 2005, \apjl, 623, L109
\bibitem[Osterbrock \& Ferland(2006)]{ost06} Osterbrock D. \& Ferland G., Astrophysics of Gaseous Nebulae and active Galactic Nuclei, Second Edition, 2006, University Science Books, Sausalito, California.
\bibitem[Pakull \& Angebault(1986)]{pak86} Pakull, M.~W., \& Angebault, L.~P.\ 1986, \nat, 322, 511 
\bibitem[Pakull \& Mirioni(2002)]{pak02} Pakull, M., \& Mirioni, L. 2002, in ``New Visions of the X-ray Universe in the XMM-Newton and Chandra Era'', (Noordwijk: ESTEC), (astro-ph/0202488)
\bibitem[Pilyugin et al.(2004)]{pil04} Pilyugin, L.~S., V{\'{\i}}lchez, J.~M., \& Contini, T.\ 2004, \aap, 425, 849 
\bibitem[Podsiadlowski et al.(2003)]{pod03} Podsiadlowski, P., Rappaport, S., \& Han, Z.\ 2003, \mnras, 341, 385 
\bibitem[Roberts(2007)]{rob07} Roberts, T.~P.\ 2007, \apss, 311, 203 
\bibitem[Schaerer \& Stasi{\'n}ska(1999)]{scha99} Schaerer, D., \& Stasi{\'n}ska, G.\ 1999, \aap, 345, L17
\bibitem[Soria et al.(2005)]{sor05b} Soria, R., Cropper, M., \& Motch, C.\ 2005, Chinese Journal of Astronomy and Astrophysics Supplement, 5, 153 
\bibitem[Soria et al.(2006)]{sor06} Soria, R., Fender, R.~P., Hannikainen, D.~C., Read, A.~M., \& Stevens, I.~R.\ 2006, \mnras, 368, 1527 
\bibitem[Soria(2007)]{sor07} Soria, R.\ 2007, \apss, 311, 213 
\bibitem[Soria \& Kuncic(2008)]{sor08} Soria, R., \& Kuncic, Z.\ 2008, American Institute of Physics Conference Series, 1053, 103
\bibitem[Sturm et al.(2002)]{sturm02} Sturm, E., Lutz, D., Verma, A., Netzer, H., Sternberg, A., Moorwood, A.~F.~M., Oliva, E., \& Genzel, R. 2002, A\&A, 393, 821
\bibitem[Zezas et al.(1999)]{zez99} Zezas, A.~L., Georgantopoulos, I., \& Ward, M.~J.\ 1999, \mnras, 308, 302


\end{thebibliography}
\end{document}